\documentstyle[aps]{revtex}

\begin{document}
\title{Hydrodynamic transport functions from quantum kinetic field theory}
\author{E. A. Calzetta \thanks{%
Electronic address: {\tt calzetta@df.uba.ar}}}
\address{Department of Physics and IAFE, University of Buenos Aires,\\
Argentina}
\author{B. L. Hu\thanks{%
Electronic address: {\tt hub@physics.umd.edu}} and S. A. Ramsey\thanks{%
Electronic address: {\tt sramsey@physics.umd.edu}}}
\address{Department of Physics, University of Maryland, College Park, Maryland 20742}
\date{January 31, 2000}
\maketitle

\begin{abstract}
Starting from the quantum kinetic field theory [E. Calzetta and B. L. Hu,
Phys. Rev. D37, 2878 (1988)] constructed from the closed-time-path (CTP),
two-particle-irreducible (2PI) effective action we show how to compute from
first principles the shear and bulk viscosity functions in the
hydrodynamic-thermodynamic regime. For a real scalar field with $\lambda
\Phi ^{4}$ self-interaction we need to include 4 loop graphs in the equation
of motion. This work provides a microscopic field-theoretical basis to the
``effective kinetic theory'' proposed by Jeon and Yaffe [S. Jeon and L. G.
Yaffe, Phys. Rev. D53, 5799 (1996)], while our result for the bulk viscosity
reproduces their expression derived from linear response theory and the
imaginary-time formalism of thermal field theory. Though unavoidably
involved in calculations of this sort, we feel that the approach using
fundamental quantum kinetic field theory is conceptually clearer and
methodically simpler than the effective kinetic theory approach, as the
success of the latter requires clever rendition of diagrammatic resummations
which is neither straightforward nor failsafe. Moreover, the method based on
the CTP-2PI effective action illustrated here for a scalar field can be
formulated entirely in terms of functional integral quantization, which
makes it an appealing method for a first-principles calculation of transport
functions of a thermal non-abelian gauge theory, e.g., QCD quark-gluon
plasma produced from heavy ion collisions.
\end{abstract}

\newpage

\section{Introduction}

In a recent series of papers \cite{jeon:1993a,jeon:1995a,jeon:1996a}, Jeon
and Yaffe (JY) have derived expressions for the transport functions for a
real, self-interacting scalar field in flat space, from first principles,
using the Kubo formulas \cite{kadanoff:1963a} and the imaginary-time
formalism of thermal field theory. A necessary step in evaluating transport
functions using the Kubo formulas is taking the zero-frequency limit of the
time-Fourier-transformed spatial correlator of the energy momentum tensor.
In order to avoid infrared divergences which arise in taking the $\omega
\rightarrow 0$ limit, a complicated set of resummations of ladder diagrams
had to be performed \cite{jeon:1995a}. However, it was observed in \cite
{jeon:1996a} that the same integral equations for the transport functions
obtained from the Kubo formulas can be derived from an {\it effective
kinetic theory}{\em \/} which takes into account the one-loop
finite-temperature corrections to the mass of quasiparticles in the scalar
theory, and to the effective vertices for quasiparticle scattering. In their
treatment, the effective kinetic theory is presented as a physically
well-motivated, but phenomenological, theory, whose justification is taken
to be the fact that it gives transport properties which agree with the
leading-order nonperturbative results computed using thermal field theory.

We demonstrate in this work that, in fact, the effective kinetic equations
for $\lambda \Phi ^4$ theory proposed by Jeon and Yaffe are derivable from a
kinetic theory of quantum fields constructed earlier by two of us \cite
{calzetta:1988b} following the work of Kadanoff and Baym et al \cite
{kadanoff:1962} and continued by many others \cite{daniel:1984}. In
particular their effective kinetic equations are derivable in our approach
from the closed-time-path (CTP), two-particle-irreducible (2PI) effective
action \cite{calzetta:1988b} truncated at four loops for the $\lambda \Phi ^4
$ theory. Even though the calculation of higher loop effects is necessarily
involved technically, we feel the quantum kinetic field theory approach is
conceptually clearer and methodically simpler than Jeon and Yaffe's
approach, as the success of the latter requires clever rendition of
diagrammatic resummations which is neither straightforward nor fail-safe.
For example, the usual form of Boltzmann equation derived in, say, \cite
{calzetta:1988b} assumes 2-2 particle scattering, which conserves particle
number, but bulk viscosity arises from particle nonconserving processes
(this was emphasized by JY). When we take into account 2-4 or 4-2 processes,
we need to go to 4 loop diagrams in the Boltzmann and the gap equations
which will no longer assume the simpler form familiar in the usual
derivations based on 2-2 processes. This extra effort is expected for
tackling higher order effects but within the same context of the same
fundamental (not effective) kinetic theory. We show that the derivation of
the effective kinetic equations from the CTP-2PI effective action requires
only the basic assumptions of kinetic theory: namely, the existence of a
separation of macroscopic and microscopic time scales \footnote{%
This means that the short wavelenght quantum fluctuations of the field are
in a state of near {\it local} thermal equilibrium, whose properties vary
slowly on the scale of the correlation length for the field. We also assume
a nearly Gaussian initial state, so that low order correlation functions are
sufficient to capture the dominant physical processes.} and the existence of
well-defined (perhaps modified) `on-shell' asymptotic particle states \cite
{lifshitz:1981a}. The assumption of weak coupling is also necessary in order
to justify the neglect of yet higher-order scattering terms which arise in
the collision integral for the derivation of a generalized Boltzmann
equation.

The apparent reported `failure' of existing kinetic equations in the
literature to account for the bulk viscosity of the scalar quantum field is
due to the fact that previous work along this line entails perturbative
expansions to insufficient accuracy, and not to any flaw or defect of the
theory. When the calculation to a sufficient accuracy is performed, as done
here, fundamental quantum kinetic field theory produces the ``effective
kinetic theory'' of Jeon and Yaffe, in particular the result for the bulk
viscosity reproduces JY's expression derived from linear response theory.
More importantly perhaps, our method based on the CTP-2PI effective action
illustrated here for a scalar field can be formulated entirely in terms of
functional integral quantization, which makes it an appealing method for a
first-principles calculation of transport functions of a thermal non-abelian
gauge theory, e.g., QCD quark-gluon plasma produced from heavy ion
collisions. First- principles approach with a clear bearing on fundamental
physics, involving nonlinear and nonperturbative effects, such as those
employed here for this task, and elsewhere for related problems, are, in our
opinion, essential for the successful establishment of a viable and useful
quantum field theory of non-equilibrium processes. (For a sampling of
current activities, see, e.g., \cite{curapp}.)

\subsection{Statement of the problem}

Let us begin by stating in simple terms what are the transport coefficients
to be computed \cite{llfd,hydro,israel88}. For a real scalar field, there
are no other conserved quantities than energy and momentum. Since there is
no fundamental concept of particle number (equivalently, since the field
describes both particles and antiparticles), chemical potential must be set
to zero identically in the grand canonical (equilibrium) density matrix.
Equilibrium states are parametrized by the vector $\beta ^{\mu }=\beta
u^{\mu }$,where $u^{\mu }$ is a timelike unit vector ($u^{\mu }u_{\mu }=-1$)
and $\beta =1/T$ is the inverse temperature.

In the semiclassical limit, we define the energy momentum tensor $T^{\mu \nu
}$ as the expectation value of the corresponding Heisenberg operator. $%
T^{\mu \nu }$ may be decomposed with respect to $u^\mu $ as

\begin{equation}
T^{\mu \nu }=\rho u^\mu u^\nu +p P ^{\mu \nu };\qquad P ^{\mu \nu }=\eta
^{\mu \nu }+u^\mu u^\nu :\qquad P ^{\mu \nu }u_\nu =0  \label{def1}
\end{equation}
where $\eta ^{\mu \nu }$ is the Minkowsky metric diag (-1,1,1,1) (for
simplicity we shall work in flat spacetime, although a generally covariant
formulation is readily available), thus defining the energy density $\rho $
and pressure $p$. Observe that there is no {\it heat conduction}, namely,
that in the rest frame there is no energy flux. In this case this shall be
true even in nonequilibrium states, since there are no currents other than
four - velocity to break the isotropy of space in the rest frame. In other
words, we are forced to adopt the Landau - Lifshitz formulation over
Eckart's. (See footnote below.)

The functional dependences $\rho =\rho \left( T\right) $ and $p=p\left(
T\right) $ define the equation of state in parametric form. The macroscopic
description is completed by giving a concrete expression for the entropy
flux $S^\mu $, which must be consistent with the first law of thermodynamics 
$dS^\mu =-\beta _\nu dT^{\mu \nu }$. Moreover, in equilibrium, vanishing
entropy production $S_{,\mu }^\mu =0$ must follow from energy - momentum
conservation $T_{,\nu }^{\mu \nu }=0$. If we introduce the thermodynamic
potential $\Phi ^\mu =S^\mu +\beta _\nu T^{\mu \nu }$, consistency requires $%
\Phi ^\mu =p\beta ^\mu $ , and the identity

\begin{equation}
T\frac{dp}{dT}=p+\rho  \label{thermoid}
\end{equation}
Energy - momentum conservation implies the identities (recall that $u_\mu
u_{,\nu }^\mu =0$)

\begin{equation}
\rho _{,t}-\left( \rho +p\right) u_{;\nu }^\nu =0;\qquad -\left( \rho
+p\right) u_{,t}^\mu +P ^{\mu \nu }p_{,\nu }=0  \label{edcon}
\end{equation}
where

\begin{equation}
X_{,t}\equiv -u^\mu X_{,\mu }  \label{tder}
\end{equation}

Since $\rho $ and $p$ become space dependent only through their temperature
dependence, we may write $\rho _{,t}=\rho _{,T}T_{,t}$, and similarly for $p$%
. Using the identity Eq. (\ref{thermoid}), Eqs. (\ref{edcon}) simplify to

\begin{equation}
\frac 1TT,_t-c_s^2u_{;\nu }^\nu =0;\qquad -u_{,t}^\mu +\frac 1T P ^{\mu \nu
}T_{,\nu }=0  \label{sound}
\end{equation}
where $c_s=\sqrt{p_{,T}/\rho _{,T}}$ is the speed of sound.

Let us now consider a near equilibrium state, meaning that the properties of
the actual state remain close to that of a conveniently chosen fiducial
equilibrium state. There is some arbitrariness in this choice, \footnote{%
There are two common choices of fiducial states. In the Eckart prescription,
one chooses an equilibrium state with the same particle current and energy
density as the actual state, and reads out the equilibrium pressure from the
equation of state. Thus in the Eckart frame there may be energy flux
relative to the particle flux, which is interpreted as heat. In the
Landau-Lifshitz prescription, the fiducial state has the same energy
density, energy flux and particle number density as the actual state. Thus
heat is read out of the particle current.} but, following the so-called
Landau - Lifshitz prescription \cite{llfd}, we shall choose the fiducial
equilibrium state as that having the same four - velocity and energy density
as the actual state. More precisely, we define the four velocity as the only
timelike unit eigenvector of the actual energy momentum tensor $T^{\mu \nu }$
(assumed to satisfy suitable energy conditions), and then the energy density
is defined as $\rho =T^{\mu \nu }u_\mu u_\nu $. Knowing $\rho ,$ we may
compute the temperature $T$ and pressure $p$ of the equilibrium state, and
thus the departure of the actual $T^{\mu \nu }$ from its value $T_0^{\mu \nu
}$ in the fiducial equilibrium state. Observe that if we write $T^{\mu \nu
}=T_0^{\mu \nu }+\delta T^{\mu \nu }$, then by construction $u_\nu \delta
T^{\mu \nu }=0.$ $\delta T^{\mu \nu }$ is usually parametrized in terms of
the bulk $\tau $ and shear $\tau ^{\mu \nu }$ stresses, as $\delta T^{\mu
\nu }=\tau P ^{\mu \nu }+\tau ^{\mu \nu }$, $\tau _\mu ^\mu =0.$

Remaining within the so-called {\it first order formalism }\cite
{israel88,fof}, we may write the entropy production in the nonequilibrium
evolution as $S_{,\mu }^{\mu }=-\beta _{\mu ,\nu }\delta T^{\mu \nu }$ (we
refer the reader to the literature for a thorough discussion of the
hypothesis involved in this formula). Decompose

\begin{eqnarray}
\beta _{\mu ,\nu } &=&-\frac 1Tu_\nu u_\mu c_s^2u_{;\nu }^\nu +\frac 1{T^2}%
\left[ P _\mu ^\lambda u_\nu T_{,\lambda }-P _\nu ^\lambda T_{,\lambda
}u_\mu \right]  \label{decomp2} \\
&&-\frac 1Tu_\nu u_\mu \left[ \frac{T_{,t}}T-c_s^2u_{;\nu }^\nu \right] +%
\frac 1T P _\mu ^\sigma u_\nu \left[ u_{\sigma ,t}-\frac 1T P _\sigma
^\lambda T_{,\lambda }\right]  \nonumber \\
&&+\frac 1T\tilde{H}_{\mu \nu }+\frac 1TH_{\mu \nu }+\frac 1{3T}P _{\mu \nu
}u_{,\rho }^\rho  \nonumber
\end{eqnarray}
where

\begin{equation}
H^{\mu \nu }=\frac 12 P ^{\mu \lambda }P ^{\nu \sigma }\left[ u_{\lambda
,\sigma }+u_{\sigma ,\lambda }-\frac 23 P _{\lambda \sigma }u_{,\rho }^\rho
\right] ;\qquad \tilde H^{\mu \nu }=\frac 12 P ^{\mu \lambda }P ^{\nu \sigma
}\left[ u_{\lambda ,\sigma }-u_{\sigma ,\lambda }\right]  \label{hmunu}
\end{equation}

The condition that entropy is created rather than destroyed leads us to
parametrize

\begin{equation}
\tau ^{\mu \nu }=-\eta H^{\mu \nu };\qquad \tau =-\zeta u_{,\rho }^\rho
;\qquad \eta ,\zeta \geq 0  \label{noneq}
\end{equation}
where $\eta $ and $\zeta $ are the {\it shear }and {\it bulk} viscosity
coefficients, respectively. We wish to compute these coefficients as
functions of the temperature and other parameters in the theory.

\subsection{Transport coefficients from kinetic equations}

Since thermodynamics alone cannot determine the dependence of the viscosity
coefficients on temperature, to proceed we must place the macroscopic
description within a more basic and comprehensive framework, i.e., kinetic
theory, where there is a well known method to extract the transport
functions \cite{kinetic,israel}.

The framework is a system described by a 1 - particle distribution function $%
f$. There is a known prescription to compute the energy momentum tensor $%
T^{\mu \nu }$ from the moments of the distribution function. In equilibrium, 
$f$ depends only on the inverse temperature 4-vector field $\beta _\mu .$
The starting point is the transport equation for $f$

\begin{equation}
\left[ p^\mu \frac \partial {\partial X^\mu }-\frac 12M_{,\mu }^2\frac 
\partial {\partial p_\mu }\right] f=I_{col}[f]  \label{transport}
\end{equation}
where $I_{col}$ is the collision integral and $M^2$ is the mass of the
particle (with possible position dependence). $f$ is assumed to be of the
form $f=f_0+\delta f$, where $f_0$ is the local equilibrium distribution

\begin{equation}
f_0=\frac 1{e^{\left| \beta _\mu ^0p^\mu \right| }-1}  \label{equilibrium}
\end{equation}
where $\beta _\mu ^0=u_\mu /T_0$ and $\delta f$ is a perturbation. Since the
collision integral vanishes identically for local thermal equilibrium, we
can write the collision integral as a linear integral operator $\hat Q$
acting on $\delta f$

\begin{equation}
I_{col}[f_0+\delta f]=\hat Q[\delta f]  \label{linearcol}
\end{equation}
On the other hand, if we neglect $\delta f$ on the left hand side of the
transport equation, we can write it as some differential operator acting on $%
\beta _\mu ^0$ , thus obtaining a linear integral equation for $\delta f$

\begin{equation}
\hat Q[\delta f]=Q_E[\partial _t,\partial _i](\beta _\mu ^0)
\label{lineartrans}
\end{equation}
The $\hat Q$ operator satisfies four constraints, which follow from energy -
momentum conservation, namely

\begin{equation}
\int \frac{d^4p}{\left( 2\pi \right) ^4}\theta \left( p^0\right) \delta
\left( \Omega _0\right) p^\mu \hat Q=0  \label{constraint}
\end{equation}
where $\Omega _0=p^2+M^2$ enforces the on-shell condition. Thus the equation
for $\delta f$ requires four integrability conditions

\begin{equation}
\int \frac{d^4p}{\left( 2\pi \right) ^4}\theta \left( p^0\right) \delta
\left( \Omega _0\right) p^\mu Q_E[\partial _t,\partial _i](\beta _\mu ^0)=0
\label{integrabcond}
\end{equation}

The integrability conditions reduce to a system of differential equations
for $\beta _\mu ^0$ , which are in fact the conservation laws for the energy
- momentum tensor $T^{\mu \nu }$ eqs. (\ref{sound}). These equations allow
us to eliminate time derivatives from the transport equation, which
simplifies to

\begin{equation}
\hat Q[\delta f]= Q_E[\partial _i](\beta _\mu ^0)  \label{reducedlintrans}
\end{equation}
On solving this equation, we determine the correction $\delta f$ to the
distribution function, and thereby the correction to the energy momentum
tensor. In general the terms containing $\delta f$ will contribute a term $%
\delta \rho $ to the energy density; thus we define the physical temperature 
$T$ from the condition $\rho \left( T\right) =\rho \left( T_0\right) +\delta
\rho $, or equivalently $T_0=T+\delta T$, where $\delta T=-\delta \rho /\rho
_{,T}.$ Knowing the temperature $T$ we may compute the energy momentum
tensor $T_0^{\mu \nu }$ in the fiducial state, subtract it from the physical 
$T^{\mu \nu }$ to determine the nonequilibrium part $\delta T^{\mu \nu }$,
and then read out the viscosity coefficients by matching it to the form
given in Eq. (\ref{noneq}).

\subsection{Transport functions in quantum kinetic field theory}

From the discussion above, we may identify the main steps involved in
computing the transport functions, namely:

1) Find a description of the system in terms of a 1-particle distribution
function $f$, and the corresponding transport equation,

2) Find the structure of equilibrium states, including the expression of
conserved currents in terms of $f$, and the equilibrium equation of state;

3) Solve the linearized transport equation to obtain the response of the
system to gradients in the hydrodynamical variables, and read out the
nonequilibrium stresses.

Step 1 is done in detail in \cite{calzetta:1988b}, where the self - energy
is computed to 2-loops accuracy, giving as a result that, for a $\lambda
\phi ^4$ type interaction, the transport equation for $f$ is simply the
relativistic Boltzmann equation for Bose particles, with the only
modification of allowing for a variable mass as in Eq. (\ref{transport}).
This Vlasov type correction takes into account the fact that the physical
mass $M^2$ of the particles is connected to the temperature through the gap
equation, and thereby $M^2$ becomes position dependent if $T$ is. The (only)
conserved current $T^{\mu \nu }$ is defined as the expectation value of the
corresponding Heisenberg operator, and the hydrodynamic variables are read
out from it, so step (2) does not present great difficulty.

The problem arises in step (3), because the Boltzmann collision operator
satisfies, besides the four conservation laws associated to energy -
momentum, a fifth constraint

\begin{equation}
\int \frac{d^{4}p}{\left( 2\pi \right) ^{4}}\theta \left( p^{0}\right)
\delta \left( \Omega _{0}\right) Q_N=0  \label{fifthconstraint}
\end{equation}
associated to the conservation of particle number in binary collisions.
There is therefore a fifth integrability condition, and the system of
macroscopic equations for $T_{0}$ and $u^{\mu }$ becomes overdeterminated.
One could hope that the fifth constraint would be true just as a consequence
of the other four, but we shall show below that in an interacting theory
this is not the case.

Continuing on this route, the linearized transport equation built out of the
Boltzmann collision operator is not integrable, and the calculation grinds
to a halt. If we are going to compute the bulk viscosity out of quantum
kinetic theory, then the collision operator cannot be just Boltzmann's
derived from 2-2 collision processes. A generalized collision operator
including particle number changing terms besides the usual binary scattering
terms is needed. Thus the fifth constraint has to be lifted to eliminate the
inconsistency. However, in the ``effective kinetic theory'' of Jeon and
Yaffe these new terms are not derived but rather postulated to match an
independent calculation of cross sections from linear response theory. We
feel that it is conceptually and methodically more gratifying if these terms
can be derived {\it ab initio} from a kinetic theory of quantum fields. This
is indeed possible, as our present work aims to demonstrate.

Since these particle number changing interactions are higher order in the
coupling constant (for pure $\lambda \phi ^4$ theory they appear at $\lambda
^4$th order), it is to be expected that they may be retrieved by simply
carrying the calculation in \cite{calzetta:1988b} to a higher loop order.
However, there appears a matter of principle: if we are going to work to
high (eventually, arbitrarily high) order in perturbation theory, we cannot 
{\it assume} that the Green functions will look anything like those of the
free theory. Thus we must first confront the need to provide a non
perturbative definition of the 1-particle distribution function, (which
should of course reduce to the one used in \cite{calzetta:1988b} at first
order in perturbation theory). In equilibrium, this problem is solved by the
Kubo - Martin - Schwinger (KMS) theorem (\cite{kms}), which implies the
proportionality of the Fourier transforms of the Hadamard and Jordan
propagators (see below). Off equilibrium, following \cite{kadanoff:1962}, we
shall define the 1-particle distribution function from the ratio of the
partial Fourier transforms of these propagators. Familiarity with the KMS
theorem and the Kadanoff - Baym equations should not blind us to the highly
nontrivial nature of this definition. With this we will then have come to a
full circle of deliberations for consistency.

For the specific goal laid out for this investigation, the main technical
difficulty lies in the analysis of the collision term giving rise to the
bulk viscosity, as it is due to the particle-changing processes which even
in the leading order already involve 4 loop self-energy diagrams. This is
one of the main tasks we need to overcome.

\subsection{Summary of the paper}

The outline above provides us with a step by step guide to computing
transport functions in quantum kinetic field theory, which we shall execute
in the following sections. As noted above, the first step is the precise
definition of the 1-particle distribution function, which is discussed in
Sec. 2. In Sec. 3 we derive the transport equation. For simplicity, after
showing that to lowest non trivial order the Boltzmann collision operator is
recovered, we shall write down only the terms related to particle number
changing interactions. Section 4 is dedicated to studying the equilibrium
states of the field, with the aim of finding the precise equation of state.
The results of Sections 3 and 4 amount to a first-principles derivation of
Jeon and Yaffe's effective kinetic theory from quantum field theory.
Finally, in Section 5 we go through the actual calculation of the bulk
viscosity, which in the appropriate limit reproduces JY's estimates from
linear response theory.

\section{Nonperturbative quantum kinetic theory}

Our specific goal is to show that by consistently extending the existing
methods of quantum kinetic field theory (see e.g., \cite{calzetta:1988b}) to
four- loop order, it is possible to account both for the shear and bulk
viscosity of an interacting scalar field, as computed by Jeon and Yaffe. We
shall consider a purely quartic interaction, although for the application to
gauge theories cubic plus quartic would seem closer to what is needed. Since
bulk viscosity entails particle number changing scattering, and these
processes appear for the first time at $O\left( \lambda ^{4}\right) $, we
must push the calculation through to five loops in the closed time path
(CTP) two - particle irreducible (2PI) effective action (EA) \cite{ctp,tpi},
which will yield four loops in the equations of motion for the propagators.
We shall assume that the background field vanishes identically, so we shall
look at the 2PI-EA as a functional of the propagators alone \cite
{cddn,stobol}.

\subsection{The model}

Let us begin with the classical action for a quartically self-interacting
scalar field in Minkowski space. Using a modification of DeWitt's notation
in which capital letters denote both spacetime ($x^\mu$) and time branch (1,
2) indices \cite{ramsey:1997a}, the action can be written

\begin{equation}
S=\frac 12\phi ^AD_{AB}\phi ^B+S_i  \label{action}
\end{equation}

\begin{equation}
D_{AB}=\left[ Z_b\Box -m_b^2\right] c_{AB};\qquad S_i[\phi ]=\frac{-1}{4!}%
\;\lambda _bc_{ABCD}\phi ^A\phi ^B\phi ^C\phi ^D,  \label{interaction}
\end{equation}
where $m_b$ is the bare ``mass'' of the field, $\lambda _b$ is the bare
coupling constant and $\phi ^A$ is the scalar field. With the benefit of
hindsight, we shall put the wave function renormalization factor $Z_b=1$,
but it should be generally included. The two- and four-index objects $c_{AB}$
and $c_{ABCD}$ are defined by their contraction into the scalar field, 
\begin{equation}
c_{AB}\phi ^A\psi ^B=\int d^{\,4}x\;\left[ \phi ^1\psi ^1-\phi ^2\psi
^2\right] \left( x\right)
\end{equation}
\begin{equation}
c_{ABCD}\phi ^A\phi ^B\phi ^C\phi ^D=\int d^{\,4}x\;\left[ \left( \phi
^1\right) ^4-\left( \phi ^2\right) ^4\right] \left( x\right)
\end{equation}
We shall use $c_{AB}$ and its inverse $c^{AB}$ to raise and lower indices,
and with the use of the Einstein convention of summing over repeated
indices, their appearance may be implicit.

We wish to derive an effective kinetic description of this theory valid at
arbitrary temperature, for sufficiently weak coupling $\lambda $, in the
case of unbroken symmetry. This assumes that the expectation value of the
Heisenberg field operator $\Phi _H$ vanishes. The two-point function $<\Phi
_H\left( x\right) \Phi _H\left( y\right) >$ is the lowest-order nonvanishing
correlation function for the space of initial conditions with which we are
presently concerned. Therefore, let us couple an external, c-number,
nonlocal source $K_{AB}$ to the scalar field as follows, 
\begin{equation}
S[\phi ]\longrightarrow S[\phi ]+\frac 12K_{AB}\phi ^A\phi ^B,
\end{equation}
and construct a quantum generating functional 
\begin{equation}
Z[K]\equiv \int D\phi \exp \left[ \frac i\hbar \left( S[\phi ]+\frac 12%
K_{AB}\phi ^A\phi ^B\right) \right]  \label{eq-dz}
\end{equation}
whose functional power series expansion contains all the $n$-point functions
of the theory. The generating functional of normalized expectation values is
given by 
\begin{equation}
W[K]\equiv -i\hbar \ln Z[K].
\end{equation}
Now, we define 
\begin{equation}
\hbar G^{AB}\equiv 2\frac{\delta W[K]}{\delta K_{AB}},  \label{variation}
\end{equation}
and construct a new functional $\Gamma [G]$ which is the Legendre transform
of $W[K]$, 
\begin{equation}
\Gamma [G]\equiv W[K]-\frac \hbar 2K_{AB}G^{AB}.
\end{equation}
It follows immediately from the above definition that 
\begin{equation}
\frac{\delta \Gamma [G]}{\delta G^{AB}}=-\frac \hbar 2K_{AB}.
\end{equation}
and $\Gamma $ obeys the integro-differential equation 
\begin{equation}
\Gamma [G]=-i\hbar \ln \left[ \int D\phi \exp \left\{ \frac i\hbar \left(
S[\phi ]-\frac 1\hbar \frac{\delta \Gamma [G]}{\delta G^{AB}}(\phi ^A\phi
^B-\hbar G^{AB})\right) \right\} \right] ,
\end{equation}

By expanding $\Gamma $ in a functional power series in $\hbar $, this
equation can be solved \cite{tpi}. The solution has the form 
\begin{equation}
\Gamma [G]=-\frac{i\hbar }2\text{Tr}\ln G+\frac{i\hbar }2{}D_{AB}G^{AB}+%
\Gamma _2[G],  \label{twopiea}
\end{equation}
where the functional $\Gamma _2$ is $-i\hbar $ times the sum of all
two-particle-irreducible diagrams with lines given by $\hbar G$ and vertices
given by the quartic interaction.

The functional $\Gamma [G]$ is the two-particle-irreducible (2PI) effective
action whose variation with respect to $G$ gives the equation of motion for
the two-point function. Because we are interested in computing transport
properties of this theory, we will need to include those terms in the
perturbative expansion for $\Gamma [G]$ which will contribute to the bulk
and shear viscosity in the weak-coupling, near-equilibrium limit. In order
to have binary scattering of quasiparticles in the effective kinetic theory,
we shall need to have a term with four propagators, which appears at $%
O(\lambda ^2)$. In order to have number-changing processes such as two
quasiparticles scattering into four (and vice versa), we need to include a
term with {\em eight\/} (six asymptotic on-shell propagators and two
internal lines \cite{jeon:1995a}), which appears at $O(\lambda ^4)$ in the
2PI effective action. This means taking into account the Feynman graphs in
Figures 1 to 5 \cite{kastening}. Taking the functional derivative with
respect to $G_{AB}$ yields a formal equation for the two-point function of
the theory,

\begin{equation}
\hbar D_{AB}-i\hbar \left( G^{-1}\right) _{AB}-\frac 12{\bf T}_{AB}+\Pi
_{AB}=0  \label{dyson}
\end{equation}
where we have singled out the tadpole term ${\bf T}$, Fig. 6. The remainder
of the self energy (which we shall refer to as the self energy, for short)
is given by the sum of the graphs in Figures 7 to 11 (Note: observe that in
the graphs Figs. 1-4 all internal lines are equivalent; in Fig. 5 we have
instead two sets of equivalent lines, marked $a$ and $b$ in the Figure. Thus
this last graph gives rise to two different graphs upon variation, i.e.,
Figs. 10 and 11). This is just the Dyson equation for the inverse
propagator, where the self energy is already expressed in terms of the
propagators themselves. In the sense of \cite{cddn,stobol} we say the self
energy has been {\it slaved} to the propagators.

There are two ways we can proceed. We can either right-multiply $G^{BC}$, or
left-multiply $G^{CA}$ into the equation, obtaining the right-multiplied and
left-multiplied Dyson equations, respectively. Note that only the tadpole
term in the is invariant under simultaneous translations of the $A$ and $B$
spacetime indices. The higher-order terms all violate translation invariance
in the equation of motion for the two-point function as a consequence of
slaving -- they describe the dissipative processes by which the system
approaches equilibrium \cite{calzetta:1988b}. From now on we shall set $%
\hbar =1$.

\subsection{Nonperturbative properties of the propagators}

Our strategy is as follows. In equilibrium, the propagators are translation
invariant, and their Fourier transform are simply proportional (KMS theorem%
\cite{kms}). Out of equilibrium, we write

\begin{equation}
G\left( x,x^{\prime }\right) =\int \frac{d^4p}{\left( 2\pi \right) ^4}%
\;e^{ipu}G\left( X,p\right)  \label{fast-slow}
\end{equation}
with $u=x-x^{\prime }$and $X=\left( x+x^{\prime }\right) /2$. We assume that 
$G(X,p)$ is slowly varying with respect to the center of mass variable $X$.

Before we start, it is useful to display the properties of the propagators
which actually follow from their definition as path ordered products of
field operators. We shall consider 8 different propagators, namely,\newline
a) the four basic propagators, appearing in equation (\ref{variation}):
Feynman $G^{11}\equiv <T\left( \Phi _{H}\left( x\right) \Phi _{H}\left(
x^{\prime }\right) \right) >$, where $T$ stands for time ordering, Dyson $%
G^{22}\equiv <\tilde{T}\left( \Phi _{H}\left( x\right) \Phi _{H}\left(
x^{\prime }\right) \right) >$, where $\tilde{T}$ stands for anti time
ordering, positive frequency $G^{21}\equiv <\Phi _{H}\left( x\right) \Phi
_{H}\left( x^{\prime }\right) >$ and negative frequency $G^{12}\equiv <\Phi
_{H}\left( x^{\prime }\right) \Phi _{H}\left( x\right) >.$The Feynman and
Dyson propagators are even. We also have

\begin{equation}
G^{11}=G^{22*};\;G^{12}=G^{21*};\;G^{12}\left( x,x^{\prime }\right)
=G^{21}\left( x^{\prime },x\right)
\end{equation}

As a consequence, $G^{11}$ and $G^{22}\left( X,p\right) $ are even functions
of momentum, while $G^{12}\left( X,p\right) =G^{21}\left( X,-p\right) .$%
Moreover, $G^{12}$and $G^{21}\left( X,p\right) $are real, and $G^{22}\left(
X,p\right) ^{*}=G^{11}$. Finally we have the identity

\begin{equation}
G^{11}+G^{22}=G^{12}+G^{21}
\end{equation}
which follows from the path ordering constraints

\begin{equation}
G^{11}=\theta \left( t-t^{\prime }\right) G^{21}+\theta \left( t^{\prime
}-t\right) G^{12}
\end{equation}

\begin{equation}
G^{22}=\theta \left( t-t^{\prime }\right) G^{12}+\theta \left( t^{\prime
}-t\right) G^{21}
\end{equation}
\newline
b) The Hadamard propagator $G_1=G^{21}+G^{12}\equiv <\left\{ \Phi _H\left(
x\right) ,\Phi _H\left( x^{\prime }\right) \right\} >$ is real and even and
therefore also is $G_1\left( X,p\right) $. The Jordan propagator $%
G=G^{21}-G^{12}\equiv <\left[ \Phi _H\left( x\right) ,\Phi _H\left(
x^{\prime }\right) \right] >$ is imaginary and odd, and so $G\left(
X,p\right) $ is odd but real. \newline
c) The advanced and retarded propagators

\begin{equation}
G_{adv}\left( x,x^{\prime }\right) =-iG\left( x,x^{\prime }\right) \theta
\left( t^{\prime }-t\right) ,\quad G_{ret}\left( x,x^{\prime }\right)
=G_{adv}\left( x^{\prime },x\right) =iG\left( x,x^{\prime }\right) \theta
\left( t-t^{\prime }\right) \;
\end{equation}
or else 
\begin{equation}
G_{ret}=i\left( G^{11}-G^{12}\right) ;\qquad G_{adv}=i\left(
G^{22}-G^{12}\right)
\end{equation}

Once $G_{ret}$ is known, we can reconstruct $G$ as

\begin{equation}
G\left( x,x^{\prime }\right) =\left( -i\right) \left[ G_{ret}\left(
x,x^{\prime }\right) -G_{ret}\left( x^{\prime },x\right) \right]
\end{equation}
So

\begin{equation}
G\left( p\right) =\left( -i\right) \left[ G_{ret}\left( p\right)
-G_{ret}\left( -p\right) \right] =2{\rm Im\;}G_{ret}\left( p\right) 
\end{equation}
where we have used that $G_{ret}\left( x,x^{\prime }\right) $is real, so $%
G_{ret}\left( -p\right) =G_{ret}\left( p\right) ^{*}.$ Also observe that $%
G_{adv}\left( p\right) =G_{ret}\left( p\right) ^{*}$

Since the retarded propagator is causal, it satisfies the equation

\begin{equation}
G_{ret}=\theta \left( t-t^{\prime }\right) G_{ret}
\end{equation}
And therefore the real and imaginary parts of its transform are Hilbert
transforms of each other

\begin{equation}
G_{ret}\left( p\right) =\frac i{2\pi }\int \frac{d\omega }{p^0-\omega
+i\varepsilon }G_{ret}\left( \omega ,\vec p\right) =\frac 12G_{ret}\left(
p\right) +\frac i{2\pi }PV\int \frac{d\omega }{p^0-\omega }G_{ret}\left(
\omega ,\vec p\right)
\end{equation}

\begin{equation}
{\rm Re}G_{ret}\left( p\right) =\frac 1\pi PV\int \frac{d\omega }{\omega -p^0%
}{\rm Im}G_{ret}\left( \omega ,\vec p\right)
\end{equation}
This implies in particular that the real and imaginary parts are orthogonal
to each other

\begin{equation}
\int d\omega \;{\rm Im}G_{ret}\left( \omega ,\vec p\right) {\rm Re}%
G_{ret}\left( \omega ,\vec p\right) =0
\end{equation}

All other propagators can be decomposed in a similar way. For example, since

\begin{equation}
G^{11}\left( x,x^{\prime }\right) =\frac 12\left[ G_1\left( x,x^{\prime
}\right) +G\left( x,x^{\prime }\right) {\rm sign}\left( t-t^{\prime }\right)
\right] =\frac 12\left[ G_1\left( x,x^{\prime }\right) -i\left(
G_{ret}\left( x,x^{\prime }\right) +G_{ret}\left( x^{\prime },x\right)
\right) \right]
\end{equation}
so

\begin{equation}
G^{11}\left( X,p\right) =\frac 12\left[ G_1-2i{\rm Re}G_{ret}\right]
;\;G^{22}\left( X,p\right) =\frac 12\left[ G_1+2i{\rm Re}G_{ret}\right]
\end{equation}

To give a nonperturbative definition of the one particle distribution
function $f$, which is the focus of attention in quantum kinetic theory, we
shall assume that the partial Fourier transforms of the Hadamard and Jordan
propagators are proportional

\begin{equation}
G_1={\rm sign}\left( p^0\right) \left[ 1+2f\right] G  \label{hadajor}
\end{equation}

Introducing a density of states function $\Delta \left( p\right) $

\begin{equation}
G\left( p\right) \equiv 2\pi {\rm sign}\left( p^0\right) \Delta \left(
p\right)
\end{equation}

then

\begin{equation}
G_1=2\pi \left[ 1+2f\right] \Delta
\end{equation}

\begin{equation}
G^{21}=2\pi \left[ \theta \left( p^0\right) +f\right] \Delta =2\pi
F^{21}\Delta
\end{equation}

\begin{equation}
G^{12}=2\pi \left[ \theta \left( -p^0\right) +f\right] \Delta =2\pi
F^{12}\Delta  \label{decomp}
\end{equation}

In equilibrium, $f$ is the Bose- Einstein distribution function (KMS
theorem). It can be assumed Eq. (\ref{hadajor}) serves as the definition of
the function $f$, valid to all orders in perturbation theory. Observe that,
since the relevant Fourier transforms are distributions (e.g., in free
theories), this definition may only be applied if both Fourier transforms
have the same singularity structure, which in last analysis is a restriction
on allowed quantum states. In what follows, we shall assume these
restrictions are met.

\subsection{The nonperturbative retarded and Jordan propagators}

In the approximation where only terms linear in the gradients of the Fourier
transforms of propagators are retained, it is possible to write down a
nonperturbative (in the coupling constant) expression for the retarded and
Jordan propagators.

Let us obtain an equation for $G_{ret}$ from, say, the equations for $G^{11}$%
and $G^{12}$, namely

\begin{equation}
G_{ret}=i\left( G^{11}-G^{12}\right) =i\left( G^{21}-G^{22}\right)
\end{equation}
leading to

\begin{equation}
-1=DG_{ret}-\frac 12{\bf T}_{11}G_{ret}+\Pi _{ret}G_{ret},
\end{equation}
where we have used that ${\bf T}_{12}=0$, and

\begin{equation}
\Pi _{ret}=\Pi _{11}+\Pi _{12}.
\end{equation}

Next we perform the Fourier transform. Since we are only interested in
computing the transport coefficients, we only need to keep terms which are
first order in gradients ( this approximation which is formally invoked in
the derivation of kinetic theory may not be always useful when dealing with
realistic physical conditions, see, e.g., \cite{mrowczinski97}). Therefore,
in computing the transforms, we shall drop all second derivative terms. The
free term $D=\Box -m_b^2$ transforms into

\begin{equation}
D=-p^2+ip\frac \partial {\partial X}+\frac 14\Box _X-m_b^2
\end{equation}
We drop the D'Alembertian as they contain second derivatives:

\begin{equation}
D\sim -p^2+ip\frac \partial {\partial X}-m_b^2
\end{equation}
The tadpole (for a generic propagator $G$) reads in position space

\begin{equation}
{\bf T}_{1B}G=\lambda G^{11}\left( x,x\right) G\left( x,x^{\prime }\right) 
\end{equation}
We write

\begin{equation}
\frac \lambda 4G_1\left( x,x\right) G\left( x,x^{\prime }\right) =\frac 
\lambda 4\int \frac{d^4p}{\left( 2\pi \right) ^4}\frac{d^4q}{\left( 2\pi
\right) ^4}e^{ipu}G_1\left( x,q\right) G\left( X,p\right)
\end{equation}
and retain only terms linear in gradients to obtain

\begin{equation}
\frac \lambda 4G_1\left( x,x\right) G\left( x,x^{\prime }\right) =\frac 
\lambda 4\int \frac{d^4p}{\left( 2\pi \right) ^4}\frac{d^4q}{\left( 2\pi
\right) ^4}e^{ipu}G\left( X,p\right) \left[ G_1\left( X,q\right) +\frac u2%
\frac \partial {\partial X}G_1\left( X,q\right) \right]
\end{equation}

The contribution to the equation has the form

\begin{equation}
\left[ -\delta M^2\left( X\right) -\frac i2\frac{\partial \left( \delta
M^2\right) }{\partial X}\frac \partial {\partial p}\right] G\left( X,p\right)
\end{equation}
where

\begin{equation}
\delta M^2\left( X\right) =\frac \lambda 4\int \frac{d^4q}{\left( 2\pi
\right) ^4}G_1\left( x,q\right)  \label{deltam2}
\end{equation}

Let us write the remaining term collectively as

\begin{equation}
\Pi G=\int d^4y\;\Pi \left( x,y\right) G\left( y,x^{\prime }\right)
\end{equation}
which transforms into

\begin{equation}
\int d^4y\int \frac{d^4p}{\left( 2\pi \right) ^4}\frac{d^4q}{\left( 2\pi
\right) ^4}\;e^{ip\left( y-x^{\prime }\right) }e^{iq\left( x-y\right) }\Pi
\left( \frac{x+y}2,q\right) G\left( \frac{x^{\prime }+y}2,p\right)
\end{equation}
Keeping only first terms in gradients, this transforms into

\begin{eqnarray}
&&\int d^4y\int \frac{d^4p}{\left( 2\pi \right) ^4}\frac{d^4q}{\left( 2\pi
\right) ^4}\;e^{ip\left( y-x^{\prime }\right) }e^{iq\left( x-y\right) } \\
&&\left\{ \Pi \left( X,q\right) G\left( X,p\right) +\frac{y-x^{\prime }}2%
\left( \partial _X\Pi \right) \left( q\right) G\left( p\right) +\frac{y-x}2%
\Pi \left( q\right) \partial _XG\left( p\right) \right\}  \nonumber
\end{eqnarray}
and then into

\begin{eqnarray}
&&\int d^4y\int \frac{d^4p}{\left( 2\pi \right) ^4}\frac{d^4q}{\left( 2\pi
\right) ^4}\;e^{ip\left( y-x^{\prime }\right) }e^{iq\left( x-y\right) } \\
&&\left\{ \Pi \left( X,q\right) G\left( X,p\right) +\frac i2\left( \partial
_X\Pi \right) \left( q\right) \partial _pG\left( p\right) -\frac i2\partial
_q\Pi \left( q\right) \partial _XG\left( p\right) \right\}  \nonumber
\end{eqnarray}
which contributes a term

\begin{equation}
\left[ \frac i2\left( \left( \partial _X\Pi _{ret}\right) \partial _p-\left(
\partial _p\Pi _{ret}\right) \partial _X\right) +\Pi _{ret}\left( p\right)
\right] G_{ret}
\end{equation}
to the equation of motion.

Introducing the Poisson bracket

\begin{equation}
\left\{ f,g\right\} =\partial _pf\partial _Xg-\partial _Xf\partial _pg
\end{equation}
we may write the equation for $G_{ret}$ as

\begin{equation}
-1=-\Omega G_{ret}+\frac i2\left\{ \Omega ,G_{ret}\right\}
\end{equation}
where

\begin{equation}
\Omega =p^2+M^2-\Pi _{ret}\left( p\right) ,\qquad M^2=m_b^2+\delta M^2
\label{phiym}
\end{equation}
and we get the {\it exact} (formal) solution

\begin{equation}
G_{ret}=\left[ Z_b\left( p+i\varepsilon \right) ^2+M^2-\Pi _{ret}\left(
p\right) \right] ^{-1}=\left. \frac 1\Omega \right| _{Imp^0\rightarrow 0^{+}}
\end{equation}
where

\begin{equation}
\left( p+i\varepsilon \right) ^2=-\left( p^0+i\varepsilon \right) ^2+\vec p^2
\end{equation}
(we have displaced $p^0$ into the complex plane to account for the retarded
boundary conditions). Now write

\begin{equation}
G_{ret}={\rm Re}G_{ret}+\frac i2G
\end{equation}
Then

\begin{equation}
G=\frac{-2Im\Omega }{\left| \Omega \right| ^2};\qquad \Delta =\frac{-{\rm %
sign}\left( p^0\right) Im\Omega }{\pi \left| \Omega \right| ^2}
\label{rhoimphi}
\end{equation}

\subsection{Equation for the negative frequency propagator}

The equation for the negative frequency propagator is

\begin{equation}
DG^{12}-\frac 12{\bf T}_{1B}G^{B2}+\Pi _{1B}G^{B2}=0
\end{equation}

Recall that

\begin{equation}
G^{22}=G^{12}+iG_{adv},\qquad G_{ret}^{*}=\frac 1{\Omega ^{*}}=\frac \Omega {%
\left| \Omega \right| ^2}
\end{equation}

After Fourier transforming, we obtain

\begin{equation}
0=-\Omega \left[ G^{12}-\frac{i\Pi _{12}}{\left| \Omega \right| ^2}\right] +%
\frac i2\left\{ \Omega ,G^{12}\right\} +\frac 1{2\Omega ^{*2}}\left\{ \Omega
^{*},\Pi _{12}\right\}
\end{equation}

In keeping with the stipulation to consider only first order corrections to
local thermal equilibrium, we shall neglect all terms containing both
derivatives and radiative corrections. So the equation is equivalent to

\begin{equation}
0=-\Omega \left[ G^{12}-\frac{i\Pi _{12}}{\left| \Omega \right| ^2}\right] +%
\frac i2\left\{ \Omega ,G^{12}\right\}  \label{two}
\end{equation}
To separate this equation into real and imaginary parts, we must notice that
the combination $i\Pi _{12}$ is actually real (see Appendix).

\subsection{The unperturbed theory}

The unperturbed theory is obtained by neglecting the $O\left( \lambda
^2\right) $ terms in our equations. The unperturbed equations are

\begin{equation}
\Omega _0=p^2+M^2  \label{phiunp}
\end{equation}
and

\begin{equation}
\Delta (p)=\delta \left( p^2+M^2\right) +O\left( \lambda ^2\right)
\end{equation}
These are exact solutions of the above equations.

Concerning the distribution function, the real part of Eq. (\ref{two}) above
shows that $G^{12}$ is concentrated on the zeroes of $\Omega $, as required
by eq. (\ref{hadajor}), and, since $\left\{ \Omega _0,\Delta \right\} =0,$%
the imaginary part becomes the unperturbed transport equation

\[
0=\Delta \left[ p\frac \partial {\partial X}-\frac 12\partial _XM^2\partial
_p\right] F^{12} 
\]
which is in the form of a Vlasov equation.

\section{The transport equation}

The nonperturbative (in the coupling constant) equation we have derived for $%
G^{12}$, plus the decomposition eq. (\ref{decomp}) lead in a straightforward
way to the transport equation. Neglecting $\left\{ \Omega ,\Delta \right\} $
as before, we write Eq. (\ref{two}) as

\begin{equation}
0=-\Omega \left[ G^{12}-\frac{i\Pi _{12}}{\left| \Omega \right| ^2}\right]
+i\pi \Delta \left\{ {\rm Re}\Omega ,F^{12}\right\}  \label{three}
\end{equation}
Since ${\rm Im}\Omega =-\left| \Omega \right| ^2G/2$, its imaginary part
reduces to

\begin{equation}
0=\Delta \left[ -F^{12}{\rm Im}\Omega -\frac i2\Pi _{12}{\rm sign}\left(
p^0\right) +\frac 12\left\{ {\rm Re}\Omega ,F^{12}\right\} \right]
\label{four}
\end{equation}
which is the Boltzmann equation. To simplify it even further, we observe
that since ${\rm sign}\left( p^0\right) =F^{21}-F^{12}$, 
\begin{equation}
{\rm Im}\Omega =-{\rm Im}\Pi _{ret}=-{\rm Im}\Pi _{11}+i\Pi _{12}=(i/2)(\Pi
_{12}-\Pi _{21})  \label{imaphi}
\end{equation}
(see Appendix), so

\begin{equation}
0=\Delta \left[ \frac 12\left\{ {\rm Re}\Omega ,F^{12}\right\} -\frac i2%
\left( \Pi _{12}F^{21}-\Pi _{21}F^{12}\right) \right]  \label{fourb}
\end{equation}

This equation is formally valid to all orders in the coupling constant.
However, it is convenient to consider the loop expansion of $\Pi $ to reduce
this equation to a more familiar form.

\subsection{The collision term}

In this subsection we shall consider the expansion of the self energy $\Pi $
in terms of Feynman graphs of increasing loop order, as a means to obtain a
definite expression for the collision term in the kinetic equation (\ref
{fourb}). Since we have the relationship $\Pi _{21}\left( p\right) =\Pi
_{12}\left( -p\right) $ (see Appendix) it is enough to analyze only the
expansion of $\Pi _{12}.$ Physically this means considering only the gain
processes, which produce a particle within a given phase space cell. The
collision term is then obtained by subtracting the loss processes, which
remove a particle therein.

The first term in the expansion is the single two loop graph Fig. 7. To this
order,

\begin{equation}
\Pi _{12}\left( x,y\right) =\frac{-i}6\lambda ^2\Sigma ^{12}\left(
x,y\right) =\frac{-i}6\lambda ^2\left[ G^{12}\left( x,y\right) \right] ^3
\label{twolooppi}
\end{equation}
In momentum space, dealing with the propagators as if they were translation
invariant,

\begin{equation}
\Sigma ^{12}\left( p\right) =\left( 2\pi \right) ^4\int \frac{d^4r}{\left(
2\pi \right) ^4}\frac{d^4s}{\left( 2\pi \right) ^4}\frac{d^4t}{\left( 2\pi
\right) ^4}\delta \left( p-r-s-t\right) G^{12}\left( r\right) G^{12}\left(
s\right) G^{12}\left( t\right),
\end{equation}
and, using the definition eq. (\ref{decomp}), we get

\begin{equation}
\Sigma ^{12}\left( p\right) =\left( 2\pi \right) ^4\int \frac{d^4r\Delta
\left( r\right) }{\left( 2\pi \right) ^3}\frac{d^4s\Delta \left( s\right) }{%
\left( 2\pi \right) ^3}\frac{d^4t\Delta \left( t\right) }{\left( 2\pi
\right) ^3}\delta \left( p-r-s-t\right) F^{12}\left( r\right) F^{12}\left(
s\right) F^{12}\left( t\right)  \label{twoloopsig}
\end{equation}

If we substitute $\Delta $ by its lowest order value $\Delta_0 =\delta
\left( p^2+M^2\right) $, this yields the collision term given earlier in 
\cite{calzetta:1988b}. This represents binary collisions, which conserves
particle number. For the reasons discussed in the Introduction, it leads to
an inconsistency when one tries to compute the bulk viscosity coefficient.

The first correction to Eq. (\ref{twoloopsig}), within the two- loop theory,
comes from the radiative corrections to the density of states, as given by
eqs. (\ref{rhoimphi}) and (\ref{imaphi}). We write

\begin{equation}
\Sigma ^{12}=\Sigma _0^{12}+\delta \Sigma ^{12}
\end{equation}
where $\Sigma _0^{12}$ is the lowest order result just discussed, and (
writing $\Omega _{0s}=s^2+M^2$ for short)

\begin{eqnarray}
\delta \Sigma ^{12}\left( p\right) &=&3i\left( 2\pi \right) ^3\int \frac{d^4r%
{\rm sign}\left( r^0\right) }{\left( 2\pi \right) ^3}\frac{d^4s\delta \left(
\Omega _{0s}\right) }{\left( 2\pi \right) ^3}\frac{d^4t\delta \left( \Omega
_{0t}\right) }{\left( 2\pi \right) ^3}\delta \left( p-r-s-t\right)
\label{twoloopdelsig} \\
&&G_{ret}\left( r\right) G_{adv}\left( r\right) F^{12}\left( r\right)
F^{12}\left( s\right) F^{12}\left( t\right) \left[ \Pi _{21}-\Pi
_{12}\right] \left( r\right)  \nonumber
\end{eqnarray}

We use again eq. (\ref{twolooppi}) to get

\begin{eqnarray}
\delta \Sigma ^{12}\left( p\right) &=&\frac{-\lambda ^2\left( 2\pi \right) ^4%
}2\int \frac{d^4s\delta \left( \Omega _{0s}\right) }{\left( 2\pi \right) ^3}%
\frac{d^4t\delta \left( \Omega _{0t}\right) }{\left( 2\pi \right) ^3}\frac{%
d^4u\delta \left( \Omega _{0u}\right) }{\left( 2\pi \right) ^3}\frac{%
d^4v\delta \left( \Omega _{0v}\right) }{\left( 2\pi \right) ^3}\frac{%
d^4w\delta \left( \Omega _{0w}\right) }{\left( 2\pi \right) ^3}  \label{tlb}
\\
&&\delta \left( p+u+v+w+s+t\right) \sigma \left( u+v+w\right)  \nonumber \\
&&F^{21}\left( s\right) F^{21}\left( t\right) \left[ F^{12}\left( u\right)
F^{12}\left( v\right) F^{12}\left( w\right) -F^{21}\left( u\right)
F^{21}\left( v\right) F^{21}\left( w\right) \right]  \nonumber
\end{eqnarray}
where

\[
\sigma \left( r\right) ={\rm sign}\left( r^0\right) G_{ret}\left( r\right)
G_{adv}\left( r\right) F^{21}\left( r\right) 
\]

A successful contribution to the gain part of the collision term describing
scattering of two into four particles (this being the simplest particle
number non conserving process in this theory) must involve, besides the
factor $(1+f_p)$ already explicit in eq. (\ref{fourb}), five other factors $%
f $ or $1+f$ evaluated on on-shell momenta adding up to $p$. But equation (%
\ref{tlb}) cannot contain a term like this, because of the interference
between the two terms in brackets. After all cancellations, we are left with
radiative corrections to the already known binary collision term. We
conclude that to order $\lambda ^4$ there are no contributions to a particle
number nonconserving collision term arising from the setting sun graph. We
must consider instead the higher loop graphs, figs. 8 to 11.

\subsection{Higher loops}

Generally speaking, we expect the collision term to describe both particle
number conserving ($2\rightarrow 2$) and changing ($2\rightarrow 4$)
scattering. Because of parity, we do not expect transitions between an even
and odd number of particles. The $2\rightarrow 2$ scattering is already
present in the two-loop theory, and any further correction to it will not
contribute to the transport functions. So from the three and four loop
contributions we shall seek only terms related to $2\rightarrow 4$
scattering.

Since we only seek the lowest order contribution to the bulk viscosity, we
may substitute the density of states $\Delta $ by a delta function
concentrated on mass shell, so the notions of on and off shell recover their
usual meaning. It is then possible to ascertain from the momentum flow in
the graph whether the condition of five on-shell momenta adding to $p$ may
be fulfilled: this is just the question of whether it is possible to cut the
graph by going across five internal lines \cite{cutrules}. The three loop
contribution cannot satisfy this criterium, and we shall not analyze it
further (it only renormalizes the binary scattering amplitude). For the same
reason, we discard the graph in Fig 9, and concentrate on the graphs in
Figs. 10 and 11, which {\it a priori }pass the test.

The complete contribution to $\Pi _{12}$ from the graph in Fig. 10 reads

\begin{eqnarray}
&&\frac{-i\lambda ^4}4\left( 2\pi \right) ^{12}\int \frac{d^4q}{\left( 2\pi
\right) ^4}\frac{d^4r}{\left( 2\pi \right) ^4}\frac{d^4s}{\left( 2\pi
\right) ^4}\frac{d^4t}{\left( 2\pi \right) ^4}\frac{d^4u}{\left( 2\pi
\right) ^4}\frac{d^4v}{\left( 2\pi \right) ^4}\frac{d^4w}{\left( 2\pi
\right) ^4}  \nonumber \\
&&\delta \left( q+r+s-p\right) \delta \left( t+u+v-q\right) \delta \left(
u+t+r-w\right)  \nonumber \\
&&\left\{ 2G^{11}\left( q\right) G^{12}\left( r\right) G^{12}\left( s\right)
G^{12}\left( t\right) G^{12}\left( u\right) G^{12}\left( v\right)
G^{22}\left( w\right) \right.  \nonumber \\
&&-G^{11}\left( q\right) G^{11}\left( r\right) G^{12}\left( s\right)
G^{11}\left( t\right) G^{11}\left( u\right) G^{12}\left( v\right)
G^{12}\left( w\right)  \nonumber \\
&&\left. -G^{12}\left( q\right) G^{12}\left( r\right) G^{12}\left( s\right)
G^{22}\left( t\right) G^{22}\left( u\right) G^{22}\left( v\right)
G^{22}\left( w\right) \right\}  \label{figu10}
\end{eqnarray}

In a true contribution to $2\rightarrow 4$ scattering, the six on - shell
momenta involved (including $p$) are {\it irreducible}, in the sense that
there are no other linear relations among them than overall momentum
conservation. If we look at the three terms in curly brackets in eq. (\ref
{figu10}), we see that in the second term the three momenta $s$, $v$ and $w$
are on-shell, but they satisfy the linear relation $s+v+w-p=0$, irrespective
of the other momenta. Thus this term is not irreducible, and does not
contribute to $2\rightarrow 4$ scattering; it is another radiative
correction to the binary collision term. The same analysis disposes of the
third term, since here the on-shell momenta $q$, $r$ and $s$ are constrained
to satisfy $q+r+s-p=0.$ We will disregard these two terms.

The graph in Fig. 11 contributes

\begin{eqnarray}
&&\ \frac{-i\lambda ^4}4\left( 2\pi \right) ^{12}\int \frac{d^4q}{\left(
2\pi \right) ^4}\frac{d^4r}{\left( 2\pi \right) ^4}\frac{d^4s}{\left( 2\pi
\right) ^4}\frac{d^4t}{\left( 2\pi \right) ^4}\frac{d^4u}{\left( 2\pi
\right) ^4}\frac{d^4v}{\left( 2\pi \right) ^4}\frac{d^4w}{\left( 2\pi
\right) ^4}  \nonumber \\
&&\ \delta \left( q+r+s-p\right) \delta \left( t+u+v-q\right) \delta \left(
t+r+s-w\right)  \nonumber \\
&&\ \left\{ G^{11}\left( q\right) G^{12}\left( r\right) G^{12}\left(
s\right) G^{12}\left( t\right) G^{12}\left( u\right) G^{12}\left( v\right)
G^{22}\left( w\right) \right.  \nonumber \\
&&+G^{12}\left( q\right) G^{11}\left( r\right) G^{11}\left( s\right)
G^{21}\left( t\right) G^{22}\left( u\right) G^{22}\left( v\right)
G^{12}\left( w\right)  \nonumber \\
&&\ -G^{11}\left( q\right) G^{11}\left( r\right) G^{11}\left( s\right)
G^{11}\left( t\right) G^{12}\left( u\right) G^{12}\left( v\right)
G^{12}\left( w\right)  \nonumber \\
&&\ \left. -G^{12}\left( q\right) G^{12}\left( r\right) G^{12}\left(
s\right) G^{22}\left( t\right) G^{22}\left( u\right) G^{22}\left( v\right)
G^{22}\left( w\right) \right\}  \label{figu11}
\end{eqnarray}
Only the first term in curly brackets is irreducible. Retaining only the
irreducible contributions from both graphs, we get the prospective particle
number nonconserving collision term as

\begin{eqnarray*}
&&\ \frac{-i\lambda ^4}4\left( 2\pi \right) ^4\int \frac{d^4r}{\left( 2\pi
\right) ^4}\frac{d^4s}{\left( 2\pi \right) ^4}\frac{d^4t}{\left( 2\pi
\right) ^4}\frac{d^4u}{\left( 2\pi \right) ^4}\frac{d^4v}{\left( 2\pi
\right) ^4}\delta \left( r+s+t+u+v-p\right) \\
&&\ \sigma ^2\left( -p,r,s,t,u,v\right) G^{12}\left( r\right) G^{12}\left(
s\right) G^{12}\left( t\right) G^{12}\left( u\right) G^{12}\left( v\right)
\end{eqnarray*}

\begin{equation}
\sigma ^2=G^{11}\left( p+r+s\right) \left[ 2G^{22}\left( p+v+s\right)
+G^{22}\left( p+u+v\right) \right]  \label{coliterm}
\end{equation}
It is clear that only the totally symmetric (as a function of $r$, $s$, $t$, 
$u$ and $v$) part $\sigma _s^2$ of $\sigma ^2$ contributes to the integral,
so we shall assume that $\sigma ^2$ has been symmetrized.

To reduce eq. (\ref{coliterm}) to a more familiar form, let us assume that $%
p^0>0,$ and restrict the integral to future oriented momenta (that is, when
a momentum is past oriented, we reverse its sign). Because of momentum
conservation, some momenta must be future oriented, but because they are all
on mass-shell, they cannot be all future oriented at the same time; the
number of future oriented momenta can only be $4$, $3$ or $2$. The terms
with three future oriented momenta describe $3\rightarrow 3$ scattering,
which conserves particle number, so they are not related to the bulk
viscosity. With these considerations we finally get the particle number
nonconserving collision term as

\begin{eqnarray}
I_{2\rightarrow 4} &=&\int \frac{d^4r\theta \left( r^0\right) \delta \left(
\Omega \right) }{\left( 2\pi \right) ^3}\frac{d^4s\theta \left( s^0\right)
\delta \left( \Omega \right) }{\left( 2\pi \right) ^3}\frac{d^4t\theta
\left( t^0\right) \delta \left( \Omega \right) }{\left( 2\pi \right) ^3}%
\frac{d^4u\theta \left( u^0\right) \delta \left( \Omega \right) }{\left(
2\pi \right) ^3}\frac{d^4v\theta \left( v^0\right) \delta \left( \Omega
\right) }{\left( 2\pi \right) ^3}  \label{i24} \\
&&\ \ \ \left\{R_1 \delta_1 \left[ \left( 1+f_p\right) \left( 1+f_r\right)
\left( 1+f_s\right) \left( 1+f_t\right) f_uf_v-\left( 1+f_u\right) \left(
1+f_v\right) f_pf_rf_sf_t\right] \right.  \nonumber \\
&&\ \ \ \left. +R_2 \delta_2 \left[ \left( 1+f_p\right) \left( 1+f_r\right)
f_sf_tf_uf_v-\left( 1+f_s\right) \left( 1+f_t\right) \left( 1+f_u\right)
\left( 1+f_v\right) f_pf_r\right] \right\}  \nonumber
\end{eqnarray}
where

\begin{equation}
R_1\equiv \frac{5\lambda ^4}4\left( 2\pi \right) ^4\sigma _s^2\left(
-p,-r,-s,-t,u,v\right) ;\qquad R_2\equiv \frac{5\lambda ^4}8\left( 2\pi
\right) ^4\sigma _s^2\left( -p,-r,s,t,u,v\right)   \label{r12}
\end{equation}
\begin{equation}
\delta _1\equiv \delta \left( p+r+s+t-u-v\right) ;\qquad \delta _2\equiv
\delta \left( p+r-s-t-u-v\right) .  \label{d12}
\end{equation}

\section{Thermodynamics from quantum kinetic theory}

Our goal in this Section is to investigate the thermodynamic and
hydrodynamic properties of a quantum field, particularly the equation of
state and the speed of sound. Our starting point is the on - shell Boltzmann
equation (\ref{four}). To render the Poisson bracket manageable, we keep
only the unperturbed $\Omega $, equation (\ref{phiunp}), where $M^{2}$ is
given self consistently by Eqs (\ref{deltam2}) and (\ref{phiym}), namely

\begin{equation}
M^2=m_b^2+\delta M^2;\qquad \delta M^2\left( X\right) =\frac{\lambda _b}2%
\int \frac{d^4p}{\left( 2\pi \right) ^3}\delta \left( \Omega _0\right)
\left[ \frac 12+f\left( X,p\right) \right]  \label{gap}
\end{equation}

The kinetic equation can be written as

\begin{equation}
\frac 12\left\{ \Omega _0,f\right\} =I_{col}\left( X,p\right)
\end{equation}
where $I$ satisfies the constraint

\begin{equation}
\int \frac{d^4p}{\left( 2\pi \right) ^4}\theta \left( p^0\right) \delta
\left( \Omega _0\right) p^\mu I_{col}\left( X,p\right) =0
\end{equation}
which expresses energy - momentum conservation. Our concern is to
investigate this (only) conservation law, but first, we need to express the
gap equation eq. (\ref{gap}) in terms of finite quantities.

\subsection{The gap equation}

Let us write the gap equation as

\begin{equation}
M^2=m_b^2+m_V^2+\frac{\lambda _b}2M_T^2
\end{equation}
where

\begin{equation}
M_T^2=\int \frac{d^4p}{\left( 2\pi \right) ^3}\delta \left( \Omega _0\right)
f\left( X,p\right)
\end{equation}

\begin{equation}
m_V^2=\frac{\lambda _b}4\int \frac{d^4p}{\left( 2\pi \right) ^3}\delta
\left( \Omega _0\right)
\end{equation}

This second quantity is actually divergent, so to evaluate it we need to
regularize it first. We shall use dimensional regularization, writing

\begin{equation}
m_V^2=\frac{\lambda _b}2\mu ^\varepsilon \int \frac{d^dp}{\left( 2\pi
\right) ^d}\frac{\left( -i\right) }{p^2+M^2-i\varepsilon }
\end{equation}
where the dimensionality $d=4-\varepsilon $. We also go to euclidean
momenta, $p^0\rightarrow ip^0$, so

\begin{equation}
m_V^2=\frac{\lambda _b}2\mu ^\varepsilon \int \frac{d^dp}{\left( 2\pi
\right) ^d}\frac 1{p^2+M^2}
\end{equation}
We obtain

\[
m_V^2=-\frac{\lambda _bM^2}{16\pi ^2}\left( \frac{M^2}{4\pi \mu ^2}\right)
^{-\varepsilon /2}\frac{\Gamma \left[ 1+\frac \varepsilon 2\right] }{%
\varepsilon \left[ 1-\frac \varepsilon 2\right] } 
\]

Write

\begin{equation}
\frac{\Gamma \left[ 1+\frac \varepsilon 2\right] }{\varepsilon \left[ 1-%
\frac \varepsilon 2\right] }=\frac 1\varepsilon +\frac 12\left( 1-\gamma
\right) +...\equiv z
\end{equation}

($\gamma =.5772...$). We get

\begin{equation}
m_V^2=-\frac{\lambda _BM^2}{16\pi ^2}\left[ z-\frac 12\ln \left( \frac{M^2}{%
4\pi \mu ^2}\right) \right].
\end{equation}

We go back to the gap equation and write it as

\begin{equation}
ZM^2=m_b^2+\frac{\lambda _b}2\left[ \frac{M^2}{16\pi ^2}\ln \left( \frac{M^2%
}{4\pi \mu ^2}\right) +M_T^2\right]
\end{equation}
where

\begin{equation}
Z=1+\frac{z\lambda _b}{16\pi ^2}
\end{equation}

We now renormalize the bare couplings

\begin{equation}
m_b^2=Zm^2;\qquad \lambda _b=Z\lambda
\end{equation}
(the actual coupling in $d$ dimensions being $\mu ^\varepsilon \lambda _b$)
to obtain the physical gap equation

\begin{equation}
M^2=m^2+\frac \lambda 2M_f^2
\end{equation}
where

\begin{equation}
M_f^2=\frac{M^2}{16\pi ^2}\ln \left( \frac{M^2}{4\pi \mu ^2}\right) +M_T^2
\end{equation}

With these we turn our attention to the energy - momentum tensor.

\subsection{Energy - momentum tensor}

To define the energy - momentum tensor, we write the effective action in a
general curved background, and then use the customary definition \cite
{birdav}

\begin{equation}
T^{\mu \nu }=\frac 2{\sqrt{-g}}\frac{\delta \Gamma }{\delta g_{\mu \nu
}^{\left( 1\right) }}
\end{equation}
where only the derivative with respect to the metric in the first time
branch is taken. The effective action itself is given by Eq. (\ref{twopiea}%
). The first term $Tr\ln G$ does not depend on the metric. Written in full,
the second term reads

\begin{equation}
\frac 12\int d^4x\;\left\{ \left. \sqrt{-g^{\left( 1\right) }}\left( \Box
_x^{\left( 1\right) }-m_b^2\right) G^{11}\left( x,x^{\prime }\right) \right|
_{x^{\prime }=x}-\left( 1\rightarrow 2\right) \right\}
\end{equation}
As usual

\begin{equation}
\frac{\delta \sqrt{-g}}{\delta g_{\mu \nu }}=\frac 12\sqrt{-g}g^{\mu \nu
};\qquad \frac{\delta g^{\mu \nu }}{\delta g_{\rho \sigma }}=-g^{\mu \rho
}g^{\nu \sigma }
\end{equation}
and so the contribution from this term to $T^{\mu \nu }$ is

\begin{equation}
\left[ -\partial ^\mu \partial ^\nu +\frac 12\eta ^{\mu \nu }\left( \Box
_x-m_b^2\right) \right] \left. G^{11}\left( x,x^{\prime }\right) \right|
_{x^{\prime }=x}
\end{equation}

In the third term, the metric appears through the $\sqrt{-g}$ factors
multiplying the coupling constants. Therefore the contribution to $T^{\mu
\nu }$takes the form

\begin{equation}
\eta ^{\mu \nu }\lambda _B\frac \delta {\delta \lambda _{1111}}\Gamma _2=-%
\frac{\lambda _b}8\eta ^{\mu \nu }\left[ G^{11}\left( x,x\right) \right]
^2+\Lambda _b\eta ^{\mu \nu }
\end{equation}
where $\Lambda _b$ contains all the higher order contributions. To the
accuracy desired, $\Lambda _b$ is position independent, and we shall not
analyze it further. Adding the two nontrivial contributions we get

\begin{equation}
T^{\mu \nu }\left( X\right) =\left. -\left[ \partial ^\mu \partial ^\nu -%
\frac 12\eta ^{\mu \nu }\Box _x\right] G^{11}\left( x,x^{\prime }\right)
\right| _{x^{\prime }=x}-\frac 12\eta ^{\mu \nu }\left[ m_b^2+\frac{\lambda
_b}4G^{11}\right] G^{11}+\Lambda _b\eta ^{\mu \nu }
\end{equation}

To write the first term in terms of the distribution function, observe that $%
\partial _{x}\rightarrow ip+\frac{1}{2}\partial _{X}$. We must neglect
second derivative terms, and observe that terms involving $p\partial _{X}$
eventually vanish because $G^{11}\left( X,p\right) $ is even in $p$. So

\begin{equation}
T^{\mu \nu }\left( X\right) =\int \frac{d^4p}{\left( 2\pi \right) ^4}%
\;\left[ p^\mu p^\nu -\frac 12\eta ^{\mu \nu }p^2\right] G^{11}\left(
X,p\right) -\frac 12\eta ^{\mu \nu }\left[ m_b^2+\frac{\lambda _b}4%
G^{11}\right] G^{11}+\Lambda _b\eta ^{\mu \nu }  \label{five}
\end{equation}
We are entitled to use the unperturbed approximation for $G^{11}$

\begin{equation}
G^{11}=\frac{\left( -i\right) }{p^2+M^2-i\varepsilon }+2\pi \delta \left(
\Omega _0\right) f\left( X,p\right)
\end{equation}
The expressions that appear in $T^{\mu \nu }$are divergent and we must
regularize them. Let us consider

\begin{equation}
T_V^{\mu \nu }=-i\int \frac{d^dp}{\left( 2\pi \right) ^d}\;\frac{\left[
p^\mu p^\nu -\frac 12\eta ^{\mu \nu }p^2\right] }{p^2+M^2-i\varepsilon }%
=\left( \frac{i\left( d-2\right) }{2d}\right) \eta ^{\mu \nu }\int \frac{d^4p%
}{\left( 2\pi \right) ^4}\;\frac{p^2}{p^2+M^2-i\varepsilon }
\end{equation}
We rotate the integral into the euclidean domain and compute the integral in 
$d=4-\varepsilon $ dimensions, so finally

\begin{equation}
T_V^{\mu \nu }=-\frac{M^4\eta ^{\mu \nu }}{32\pi ^2}\left[ z-\frac 14-\frac 1%
2\ln \left( \frac{M^2}{4\pi \mu ^2}\right) \right]
\end{equation}

We also need

\[
G^{11}\left( x,x\right) =\frac 2{\lambda _B}\delta M^2=Z^{-1}\left[ M_f^2-Z%
\frac{zm^2}{8\pi ^2}\right] 
\]
Therefore

\begin{equation}
m_b^2+\frac{\lambda _b}4G^{11}\left( x,x\right) =Z\left[ m^2+\frac \lambda {%
4Z}\left( M_f^2-Z\frac{zm^2}{8\pi ^2}\right) \right] =Z\left[ m^2+\frac 
\lambda 4\left( M_f^2-\frac{zm^2}{8\pi ^2}\right) \right] +O\left( \lambda
^2\right)
\end{equation}
and

\[
\left[ m_b^2+\frac{\lambda _b}4G^{11}\right] G^{11}=-m^2\left( \frac{zm^2}{%
8\pi ^2}-M_f^2\right) -\frac \lambda 4\left[ \left( \frac{zm^2}{8\pi ^2}%
\right) ^2+\frac{zm^2M_f^2}{4\pi ^2}-M_f^4\right] +O\left( \lambda ^2\right) 
\]
So far, we get

\begin{eqnarray}
\ \ \ \ T_V^{\mu \nu }+\frac 12\eta ^{\mu \nu }\left[ m_B^2+\frac{\lambda _B}%
4G^{11}\right] G^{11} \ &=&-\eta ^{\mu \nu }\left\{ -\frac \lambda 8\left( 
\frac{zm^2}{8\pi ^2}\right) ^2-\frac{zm^4}{32\pi ^2}-\frac{m^4}{2\lambda }%
\right.  \nonumber \\
&&\ \ \ \left. +\frac{M^2m^2}{2\lambda }-\frac{M^4}{128\pi ^2}+\frac{M^2M_T^2%
}4+O\left( \lambda ^2\right) \right\}
\end{eqnarray}

Next call

\begin{equation}
T_T^{\mu \nu }=\int \frac{d^4p}{\left( 2\pi \right) ^4}\;p^\mu p^\nu 2\pi
\delta \left( \Omega _0\right) f\left( X,p\right)
\end{equation}
and observe that

\begin{equation}
\int \frac{d^4p}{\left( 2\pi \right) ^4}\;\left[ p^\mu p^\nu -\frac 12\eta
^{\mu \nu }p^2\right] 2\pi \delta \left( \Omega _0\right) f\left( X,p\right)
=T_T^{\mu \nu }+\frac 12\eta ^{\mu \nu }M^2M_T^2
\end{equation}
so

\begin{equation}
T^{\mu \nu }=T_0^{\mu \nu }+T_f^{\mu \nu }+T_T^{\mu \nu }
\end{equation}
where

\begin{equation}
T_0^{\mu \nu }=\eta ^{\mu \nu }\left\{ \frac \lambda 8\left( \frac{zm^2}{%
8\pi ^2}\right) ^2+\frac{zm^4}{32\pi ^2}+\frac{m^4}{2\lambda }\right\}
\end{equation}

\begin{equation}
T_f^{\mu \nu }=-\Lambda _f\eta ^{\mu \nu };\qquad \Lambda _f=\frac{M^2m^2}{%
2\lambda }-\frac{M^4}{128\pi ^2}-\frac{M^2M_T^2}4+O\left( \lambda ^2\right)
\label{cosmoconst}
\end{equation}

Here, $T_{0}^{\mu \nu }$ is infinite, but state independent and conserved.
It belongs to the theory of the renormalization of the gravitational action
(see \cite{birdav}, \cite{parker} and references therein), and we shall not
consider it further.

Consistency requires that we actually neglect the $O\left( \lambda ^2\right) 
$ terms in $\Lambda _f$, or at least that we consider them as a true
(temperature independent) constant. Then we can establish the following
identity, which will be useful later on. \newline
First write

\[
M_T^2=\frac 2\lambda \left( M^2-m^2\right) -\frac{M^2}{16\pi ^2}\ln \left( 
\frac{M^2}{4\pi \mu ^2}\right) 
\]

\begin{equation}
\Lambda _f=\frac{M^2m^2}\lambda -\frac{M^4}{128\pi ^2}-\frac{M^4}{2\lambda }+%
\frac{M^4}{64\pi ^2}\ln \left( \frac{M^2}{4\pi \mu ^2}\right) +\,{\rm %
constant}
\end{equation}
Then observe that

a) $\Lambda _f$ depends on temperature only through the physical mass $M^2$,
and

b) 
\begin{equation}
\frac{d\Lambda _f}{dM^2}=\frac{m^2}\lambda -\frac{M^2}\lambda +\frac{M^4}{%
32\pi ^2}\ln \left( \frac{M^2}{4\pi \mu ^2}\right) =\frac{-1}2M_T^2
\label{identity}
\end{equation}

This is the identity we need below. This expression for the energy momentum
tensor is equivalent to that given by Jeon and Yaffe. In particular, Eq.(\ref
{identity}) implies that energy momentum conservation follows from the
transport equation.

\subsection{Entropy flux and the $H$ theorem}

Let us mention also the entropy flux

\begin{equation}
S^\mu =2\int \frac{d^4p}{\left( 2\pi \right) ^4}\theta \left( p^0\right)
\;p^\mu 2\pi \delta \left( \Omega _0\right) \left\{ \left( 1+f\right) \ln
\left( 1+f\right) -f\ln f\right\}  \label{entroflux}
\end{equation}
Associated with this, entropy generation is given by

\begin{equation}
S_{;\mu }^\mu =2\int \frac{d^4p}{\left( 2\pi \right) ^4}\;\theta \left(
p^0\right) 2\pi \delta \left( \Omega _0\right) \left[ \ln \frac{\left(
1+f\right) }f\right] I_{col}  \label{entroprod}
\end{equation}
The positivity of this integral expresses the $H$ theorem. Let us write 
\begin{equation}
I_{col}=I_{2\rightarrow 2}+I_{2\rightarrow 4}  \label{icol}
\end{equation}
where the first term is the usual binary collision term

\begin{eqnarray}
I_{2\rightarrow 2} &=&\sigma \int \frac{d^4r\theta \left( r^0\right) \delta
\left( \Omega \right) }{\left( 2\pi \right) ^3}\frac{d^4s\theta \left(
s^0\right) \delta \left( \Omega \right) }{\left( 2\pi \right) ^3}\frac{%
d^4t\theta \left( t^0\right) \delta \left( \Omega \right) }{\left( 2\pi
\right) ^3}  \label{i22} \\
&&\ \ \ \delta \left( p+r-s-t\right) \left\{ \left( 1+f_p\right) \left(
1+f_r\right) f_sf_t-\left( 1+f_s\right) \left( 1+f_t\right) f_pf_r\right\} 
\nonumber
\end{eqnarray}
and the second term involves the number changing interactions, already given
in eq. (\ref{i24}).

When inserted in Eq. (\ref{entroprod}), we find

\begin{equation}
S_{;\mu }^\mu =H_{2\rightarrow 2}+H_{2\rightarrow 4}
\end{equation}
where $H_{2\rightarrow 2}$ is the usual result \cite{kinetic}

\begin{eqnarray}
H_{2\rightarrow 2} &=&\frac 12\int \frac{d^4p\theta \left( p^0\right) \delta
\left( \Omega \right) }{\left( 2\pi \right) ^3}\frac{d^4r\theta \left(
r^0\right) \delta \left( \Omega \right) }{\left( 2\pi \right) ^3}\frac{%
d^4s\theta \left( s^0\right) \delta \left( \Omega \right) }{\left( 2\pi
\right) ^3}\frac{d^4t\theta \left( t^0\right) \delta \left( \Omega \right) }{%
\left( 2\pi \right) ^3} \\
&&\left[ \ln \frac{\left( 1+f_p\right) \left( 1+f_r\right) f_sf_t}{\left(
1+f_s\right) \left( 1+f_t\right) f_pf_r}\right] \delta \left( p+r-s-t\right)
\left\{ \left( 1+f_p\right) \left( 1+f_r\right) f_sf_t-\left( 1+f_s\right)
\left( 1+f_t\right) f_pf_r\right\}  \nonumber
\end{eqnarray}
whereas [from $I_{2 \rightarrow 4}$ in Eq. (\ref{i24}]

\begin{eqnarray}
H_{2\rightarrow 4} &=&\frac 13\int \frac{d^4p\theta \left( p^0\right) \delta
\left( \Omega \right) }{\left( 2\pi \right) ^3}\frac{d^4r\theta \left(
r^0\right) \delta \left( \Omega \right) }{\left( 2\pi \right) ^3}\frac{%
d^4s\theta \left( s^0\right) \delta \left( \Omega \right) }{\left( 2\pi
\right) ^3}\frac{d^4t\theta \left( t^0\right) \delta \left( \Omega \right) }{%
\left( 2\pi \right) ^3}\frac{d^4u\theta \left( u^0\right) \delta \left(
\Omega \right) }{\left( 2\pi \right) ^3}\frac{d^4v\theta \left( v^0\right)
\delta \left( \Omega \right) }{\left( 2\pi \right) ^3} \\
&&\left(R_1 -R_2 \right) \left[ \ln \frac{\left( 1+f_p\right) \left(
1+f_r\right) \left( 1+f_s\right) \left( 1+f_t\right) f_uf_v}{\left(
1+f_u\right) \left( 1+f_v\right) f_pf_rf_sf_t}\right] \delta \left(
p+r+s+t-u-v\right)  \nonumber \\
&&\left[ \left( 1+f_p\right) \left( 1+f_r\right) \left( 1+f_s\right) \left(
1+f_t\right) f_uf_v-\left( 1+f_u\right) \left( 1+f_v\right)
f_pf_rf_sf_t\right]  \nonumber
\end{eqnarray}
is new. Thus the $H$ theorem demands the inequality

\begin{equation}
R_1\geq R_2  \label{ineq}
\end{equation}
We expect that the integral will be dominated by grazing collisions, where
one of the reactants and one of the products carry essentially all the
momentum. In this limit, $R_1\sim 2R_2$ (see eq. (\ref{r12})), so the $H$
theorem is satisfied.

\subsection{(Local) Thermal equilibrium states}

Our next concern is to investigate the equation of state, for a local
equilibrium state described by a Planckian distribution function $f_0$ as in
Eq. (\ref{equilibrium}). The energy momentum tensor is decomposed as in Eq. (%
\ref{def1}). The thermal component $T_T^{\mu \nu }$ admits a similar
decomposition

\begin{equation}
T_{0T}^{\mu \nu }=\int \frac{d^4p}{\left( 2\pi \right) ^4}\;p^\mu p^\nu 2\pi
\delta \left( \Omega _0\right) f_0\left( X,p\right) =\rho _Tu^\mu u^\nu +p_T
P ^{\mu \nu }
\end{equation}
where 
\begin{equation}
\rho _T=\int \frac{d^4p}{\left( 2\pi \right) ^4}\;\left( up\right) ^22\pi
\delta \left( \Omega _0\right) f_0\left( X,p\right)
\end{equation}
Since $\rho _T$ and $M^2$ are scalars, we may compute them in the rest frame

\[
\rho _T=\frac 1{\pi ^2}\int_M^\infty d\omega \;\frac{\omega ^2}{e^{\beta
\omega }-1}\sqrt{\omega ^2-M^2} 
\]

\begin{equation}
M_T^2=\frac 1{\pi ^2}\int_M^\infty d\omega \;\frac 1{e^{\beta \omega }-1}%
\sqrt{\omega ^2-M^2}
\end{equation}
For the thermal pressure, we find $3p_T-\rho _T=-M^2M_T^2$, so

\begin{equation}
p_T=\frac 13\left( \rho _T-M^2M_T^2\right)
\end{equation}
The total energy density and pressure are then

\begin{equation}
\rho =\rho _T+\Lambda _f;\qquad p=p_T-\Lambda _f
\end{equation}

The equilibrium entropy flux takes the form $S_0^\mu =p\beta ^\mu -T_0^{\mu
\nu }\beta _\nu =\left( \rho +p\right) \beta ^\mu =\left( \rho _T+p_T\right)
\beta ^\mu $. On the other hand, Eq. (\ref{entroflux}) yields $S_0^\mu =\Phi
_{0T}^\mu -T_{0T}^{\mu \nu }\beta _\nu $, where

\begin{equation}
\Phi_{0T}^\mu =-2\int \frac{d^4p}{\left( 2\pi \right) ^4}\theta \left(
p^0\right) \;p^\mu 2\pi \delta \left( \Omega _0\right) \ln \left[
1-e^{-\left| \beta _\mu p^\mu \right| }\right]
\end{equation}

This form of the thermodynamic potential brings to our attention other
equivalent expressions for the thermal pressure

\begin{equation}
\frac{p_T}T=\frac{-1}{\pi ^2}\int_M^\infty d\omega \;\omega \sqrt{\omega
^2-M^2}\ln \left[ 1-e^{-\beta \omega }\right]  \label{altpressure}
\end{equation}
and

\begin{equation}
p_T=\frac 1{3\pi ^2}\int_M^\infty d\omega \;\left[ \omega ^2-M^2\right]
^{3/2}f_0  \label{intpressure}
\end{equation}

Observe that Eqs (\ref{identity}) and (\ref{altpressure}) imply the
thermodynamic relationship Eq. (\ref{thermoid}) (here and henceforth, we
shall use $d/dT$ to denote a total temperature derivative, that is with
respect to the explicit temperature dependence through $f_0$ as well as the
implicit dependence through $M^2$. We shall use $\partial /\partial T$ when
we mean only the former). Indeed, Eq. (\ref{altpressure}) implies

\begin{equation}
T\frac{dp_T}{dT}=p_T+\rho _T-\frac{M_T^2}2T\frac{dM^2}{dT}
\end{equation}
But $p_T+\rho _T=\rho +p$, and

\begin{equation}
T\frac{dp}{dT}=T\frac{dp_T}{dT}-T\frac{d\Lambda _f}{dT}
\end{equation}
So Eq. (\ref{thermoid}) follows from (\ref{identity}). This concludes our
study of the local equilibrium states

\section{Linearized transport equation}

Under local thermal equilibrium, the transport equation is violated. We have 
$I_{col}=0$, while the transport part (for $p^0>0$)

\begin{equation}
\left[ p^\mu \frac \partial {\partial X^\mu }-\frac 12M_{,\mu }^2\frac 
\partial {\partial p_\mu }\right] f_0=f_0\left( 1+f_0\right) \left[ p^\mu
p^\nu \beta _{\mu ,\nu }-\frac 12M_{,\mu }^2\beta ^\mu \right]
\end{equation}

Recalling the decomposition Eq. (\ref{decomp2}) and assuming the macroscopic
equations Eq. (\ref{sound}), the LHS of the transport equation becomes

\begin{equation}
f_0\left( 1+f_0\right) \left[ \frac 1Tp_\mu p_\nu H^{\mu \nu }-\frac 1T%
\left\{ \left( p.u\right) ^2\left[ c_s^2-\frac 13\right] +\frac{M^2}3-\frac{%
c_s^2}2TM_{,T}^2\right\} u_{,\lambda }^\lambda \right]   \label{lhs}
\end{equation}
This plays the role of the $Q_E$ differential operator in Eq. (\ref
{reducedlintrans})

\subsection{The linearized collision term}

At this point we need to shift our attention to the right hand side of the
transport equation, Eqs. (\ref{icol}), (\ref{i22}) and (\ref{i24}). The
collision term vanishes identically under local thermal equilibrium, so we
need to consider a distribution function deviating from it. Write

\begin{equation}
f=f_0+f_0\left( 1+f_0\right) \chi
\end{equation}
Since $I_{col}\left[ f_0\right] \equiv 0$, only the deviation contributes to
the collision integral. We keep only linear terms, and write, by analogy
with Eq. (\ref{lhs})

\begin{equation}
\delta I_{col}=f_{0p}\left( 1+f_{0p}\right) \left[ \delta I_{2\rightarrow
2}+\delta I_{2\rightarrow 4}\right]  \label{rhs}
\end{equation}
where, upon introducing the momentum space volume element

\begin{equation}
Dp=\frac{d^4p\theta \left( p^0\right) \delta \left( \Omega \right) }{\left(
2\pi \right) ^3}f_{0p}\left( 1+f_{0p}\right)  \label{pvol}
\end{equation}
we have 
\begin{equation}
\delta I_{2\rightarrow 2}=\sigma \int DrDsDt\;\delta \left( p+r-s-t\right) 
\frac{\left\{ -\chi _p-\chi _r+2\chi _s\right\} }{\left[ \left(
1+f_{0p}\right) \left( 1+f_{0r}\right) f_{0s}f_{0t}\right] }
\label{twobytwo}
\end{equation}
and similarly

\begin{eqnarray}
\delta I_{2\rightarrow 4} &=&\int DrDsDtDuDv  \label{twobyfour} \\
&&\ \left\{ R_1 \delta _1\frac{\left[ -\chi _p-3\chi _r+2\chi _u\right] }{%
\left[ \left( 1+f_p\right) \left( 1+f_r\right) \left( 1+f_s\right) \left(
1+f_t\right) f_uf_v\right] }+R_2 \delta _2\frac{\left[ -\chi _p-\chi
_r+4\chi _s\right] }{\left[ \left( 1+f_p\right) \left( 1+f_r\right)
f_sf_tf_uf_v\right] }\right\}  \nonumber
\end{eqnarray}
where $R_{1,2}$ and $\delta_{1,2}$ were defined in Eqs. (\ref{r12}, \ref{d12}%
).

\subsection{The method of moments}

Given Eqs (\ref{lhs}) and (\ref{rhs}), the linearized transport equation can
be rewritten as

\begin{equation}
\frac 1Tp_\mu p_\nu H^{\mu \nu }-\frac{u_{,\lambda }^\lambda }T\left\{
\left( p.u\right) ^2\left[ c_s^2-\frac 13\right] +\frac{M^2}3-\frac{c_s^2}2%
TM_{,T}^2\right\} =K\left[ \chi \right]  \label{lineareq}
\end{equation}
where $K$ is an hermitian operator in the space of functions defined on the
positive energy mass shell. We further introduce an inner product in this
space by defining

\begin{equation}
\left\langle \varsigma ,\chi \right\rangle =\int Dp\;\varsigma ^{*}\left(
p\right) \chi \left( p\right) ;\qquad \left\langle \chi \right\rangle \equiv
\left\langle 1,\chi \right\rangle  \label{inner}
\end{equation}
For our purposes, it will be enough to forfeit a rigorous solution, and to
seek instead a solution using the method of moments. This entails first
writing Eq. (\ref{lineareq}) in the orthogonal basis built out of the
monomials $1$, $p^\mu $, $p^\mu p^\nu $, etc. (always with respect to the
inner product Eq. (\ref{inner}), with $Dp$ defined as in Eq. (\ref{pvol})),
and then truncating it to only the first few moments.

To simplify our notation, let us adopt the local rest frame, and write $%
\omega =p^0=-u.p$. Let $\chi _0=1$ be the first element of our basis. The
remaining functions are ($i=1$, $2$, $3$)

\begin{eqnarray}
\chi _1 &=&\omega -\frac{\left\langle \omega \right\rangle }{\left\langle
1\right\rangle };\quad \chi _2=\omega ^2+\omega \frac{\left\langle \omega
^2\right\rangle \left\langle \omega \right\rangle -\left\langle \omega
^3\right\rangle \left\langle 1\right\rangle }{\left\langle 1\right\rangle
\left\langle \omega ^2\right\rangle -\left\langle \omega \right\rangle ^2}+%
\frac{\left\langle \omega ^3\right\rangle \left\langle \omega \right\rangle
-\left\langle \omega ^2\right\rangle ^2}{\left\langle 1\right\rangle
\left\langle \omega ^2\right\rangle -\left\langle \omega \right\rangle ^2}
\label{basis} \\
q_1^i &=&p^i;\quad q_2^i=p^i\left[ \omega -\frac{\left\langle \omega \vec{p}%
^2\right\rangle }{\left\langle \vec{p}^2\right\rangle }\right]  \nonumber
\end{eqnarray}

To this we must add five independent functions built out of the binary
products $p^ip^j$ (there are only five independent functions, because $\vec{p%
}^2=\omega ^2-M^2$ is not independent of the above). The simplest procedure
is to think of these monomials as the composition of two spin 1 objects; the
spin zero component of the composition is precisely $\vec{p}^2$, and the
spin 1 part, being antisymmetric, will vanish, so our functions are the five 
$l=2$ spherical harmonics. For example, calling $p_{\pm }=p_x\pm ip_y,$ we
may choose

\begin{equation}
Y_{m}=\left(
p_{+}^{2},p_{+}p_{z},p_{z}^{2}-p_{+}p_{-},p_{-}p_{z},p_{-}^{2}\right)
;\qquad 2\geq m\geq -2  \label{harmonics}
\end{equation}
We also have the relationships (see Appendix)

\begin{equation}
\left\langle \omega \right\rangle =T^2\frac{d\rho }{dT}\frac{\left[
1-3c_s^2\right] }{\left[ M^2-\frac 12TM_{,T}^2\right] }  \label{omega}
\end{equation}

\begin{equation}
\left\langle \omega ^3\right\rangle =T^2\frac{d\rho }{dT}\frac{\left[ M^2-%
\frac 32TM_{,T}^2c_s^2\right] }{\left[ M^2-\frac 12TM_{,T}^2\right] }
\label{omega2}
\end{equation}

\begin{equation}
\left\langle \omega \vec{p}^2\right\rangle =3c_s^2T^2\frac{d\rho }{dT}
\label{p3norm}
\end{equation}
In terms of the new functions, Eq. (\ref{lineareq}) reads

\begin{equation}
\frac 1T\Gamma _{ij}^mY_mH^{ij}+\Gamma \left\{ \left\langle \omega
\right\rangle \chi _2+c\left[ \frac{\left\langle \omega \right\rangle \chi _1%
}{\left\langle \omega ,\chi _1\right\rangle }-1\right] \right\} =K\left[
\chi \right]  \label{lineareq2}
\end{equation}
where

\begin{equation}
\Gamma =\frac{u_{,\lambda }^\lambda \left[ M^2-\frac 12TM_{,T}^2\right] }{%
3T^3\frac{d\rho }{dT}};\qquad \Gamma _{ij}^m=\frac{\left\langle Y_m,p_ip_j-%
\frac 13\vec p^2\right\rangle }{\left\langle Y_m,Y_m\right\rangle };\qquad c=%
\frac{\left\langle \omega ^3\right\rangle \left\langle 1\right\rangle
-\left\langle \omega ^2\right\rangle \left\langle \omega \right\rangle }{%
\left\langle 1\right\rangle }  \label{coeffs}
\end{equation}

When we expand the operator $K\left[ \chi \right] $ we notice that, when
truncated to the subspace spanned by the functions eq. (\ref{basis}), the
operator matrix acquires a block form, with one block corresponding to the $%
\chi _a$ functions ($a=0$, $1$ or $2$), another to the $q_a^i$, and yet
another to the $Y_m$ functions. Since there are no $q_a^i$ functions in the
left hand side of Eq. (\ref{lineareq2}), we may as well write

\begin{equation}
\chi =\frac{-1}Tb_m Y^m-\Gamma \left[ A+B\chi _1+C\chi _2\right]  \label{chi}
\end{equation}
Since $K\left[ \omega \right] =0$, the $B$ coefficient will remain
undetermined ( the left hand side of Eq. (\ref{lineareq2}) is orthogonal to $%
\omega $, so the system is integrable). We will set $B=0$ for the time
being, and postpone further discussion until we enforce the Landau -
Lifshitz conditions.

To determine the $b_m$ coefficients, we must solve the linear system

\begin{equation}
\Gamma _{ij}^mH^{ij}=X^{mn}b_n  \label{linsys}
\end{equation}
where

\begin{equation}
X^{mn}=-\frac{\left\langle Y^m,K\left[ Y^n\right] \right\rangle }{%
\left\langle Y_m,Y_m\right\rangle }  \label{matelem}
\end{equation}
By symmetry, the $X$ matrix must be diagonal

\begin{equation}
X^{mn}=b\delta ^{mn};\qquad b\geq 0  \label{israel}
\end{equation}
(for the positivity of $b,$ see Israel \cite{israel}) leading to

\begin{equation}
b^m=\frac 1b\Gamma _{ij}^mH^{ij}  \label{bcoeff}
\end{equation}

To find the $A$ and $C$ coefficients let us expand

\[
K\left[ 1\right] =\frac{\left\langle K\left[ 1\right] \right\rangle }{%
\left\langle 1\right\rangle }\left[ 1-\frac{\left\langle \omega
\right\rangle \chi _1}{\left\langle \omega ,\chi _1\right\rangle }\right]
+\beta \frac{\chi _2}{\left\langle \chi _2^2\right\rangle } 
\]

\begin{equation}
K\left[ \chi _2\right] =\frac \beta {\left\langle 1\right\rangle }\left[ 1-%
\frac{\left\langle \omega \right\rangle \chi _1}{\left\langle \omega ,\chi
_1\right\rangle }\right] +\gamma \frac{\chi _2}{\left\langle \chi
_2^2\right\rangle }
\end{equation}
where we have used $\left\langle \omega ,K\left[ \chi \right] \right\rangle
=0$. If only binary scattering is considered, then also $\left\langle
K\left[ \chi \right] \right\rangle =0$ and $\left\langle K\left[ 1\right]
\right\rangle =\beta =0$. In general, then, $\left\langle K\left[ 1\right]
\right\rangle \sim \beta \ll \gamma .$ Therefore these equations admit an
approximate solution with $C=0,$ yielding

\begin{equation}
\chi =\frac{-1}{bT}\Gamma _{ij}^mH_{ij}Y_m+c_0;\qquad c_0=\Gamma \frac{%
\left\{ \left\langle \omega ^3\right\rangle \left\langle 1\right\rangle
-\left\langle \omega ^2\right\rangle \left\langle \omega \right\rangle
\right\} }{\left| \left\langle K\left[ 1\right] \right\rangle \right| }
\label{cnot}
\end{equation}
where we have used $\left\langle K\left[ 1\right] \right\rangle \leq 0$, as
follows from the inequality Eq. (\ref{ineq}) and the identity

\begin{equation}
\left\langle K\left[ 1\right] \right\rangle =4\lambda ^4\int \frac{%
DpDrDsDtDuDv\;\left(R_2-R_1\right) \delta _1}{\left[ \left( 1+f_p\right)
\left( 1+f_r\right) \left( 1+f_s\right) \left( 1+f_t\right) f_uf_v\right] }
\end{equation}

\subsection{The temperature shift and the bulk stress}

It can be seen from Eq. (\ref{cnot}) that the correction to the distribution
function has two components. The one associated with the $H^{\mu \nu }$
tensor contributes to shear stress, but it does not induce a change in the
energy density, and therefore it is compatible with the Landau - Lifshitz
matching conditions. The constant shift of $\chi $ by $c_0$, on the other
hand, affects in principle both the energy density and the thermal mass $M_T$
. So, to enforce the Landau - Lifshitz conditions, it must be partially
compensated by a temperature shift. Concretely, if we call $T$ the
temperature of the fiducial equilibrium state, such that $\rho \left(
T\right) $ is equal to the energy density in the nonequilibrium state, then
the temperature appearing in the local equilibrium distribution function $%
f_0 $ must be $T_0=T+\delta T.$ The effect of this temperature shift is the
same as that in the coefficient $B$ in Eq. (\ref{chi}).

The distribution function and temperature shifts in turn produce a shift $%
\delta M^2$ in the physical mass, which likewise does not affect the
transport equation. However, both $\delta T$ and $\delta M^2$ are relevant
to the bulk stress. Observe that there is no shift in the four velocity $%
u^\mu .$

The three displacements $c_0,$ $\delta T$ and $\delta M^2$ are related by
the constraints that the gap equation must hold, and the total energy
density in the nonequilibrium state must be the same as in the equilibrium
state. Write the gap equation as

\begin{equation}
M^2-\varphi \left( M^2,\mu ^2\right) =\frac \lambda 2M_T^2
\end{equation}
The linearized equation then reads

\begin{equation}
\left[ 1-\varphi ^{\prime }-\frac \lambda 2\frac{\partial M_T^2}{\partial M^2%
}\right] \delta M^2=\frac \lambda 2\left[ \frac{\partial M_T^2}{\partial T}%
\delta T+c_0\left\langle 1\right\rangle \right]
\end{equation}

As a matter of fact,

\begin{equation}
\frac{\partial M_T^2}{\partial T}=\frac{\left\langle \omega \right\rangle }{%
T^2}
\end{equation}
So finally

\begin{equation}
\delta M^2=M_{,T}^2\delta T+M_{,c}^2c_0
\end{equation}
where

\begin{equation}
M_{,c}^2=T^2M_{,T}^2\frac{\left\langle 1\right\rangle }{\left\langle \omega
\right\rangle }
\end{equation}

Since the gap equation is enforced, we can look at the (cosmological)
constant $\Lambda$ as a function of $M^2$, and

\begin{equation}
\delta \Lambda _f=\frac{-1}2M_T^2\delta M^2
\end{equation}
then

\begin{equation}
\delta \rho =\rho _{,T}\delta T+\left[ \frac{\partial \rho _T}{\partial M^2}-%
\frac 12M_T^2\right] M_{,c}^2c_0+\left\langle \omega ^2\right\rangle c_0
\end{equation}

Actually

\begin{eqnarray}
\frac{\partial \rho _T}{\partial M^2}=\frac 12M_T^2-\frac{\left\langle
\omega \right\rangle }{2T}
\end{eqnarray}
so

\begin{equation}
\delta \rho =\rho _{,T}\delta T+\left[ \left\langle \omega ^2\right\rangle -%
\frac{\left\langle 1\right\rangle }2TM_{,T}^2\right] c_0.
\end{equation}
And since the total energy remains the same,

\begin{equation}
\rho _{,T}\delta T=-c_0\left[ \left\langle \omega ^2\right\rangle -\frac{%
\left\langle 1\right\rangle }2TM_{,T}^2\right] .
\end{equation}

Let us apply the same reasoning to the bulk stress, which results from both
the departure of the pressure from $p\left( T\right) $ and the direct
contribution from the new terms in the distribution function

\begin{equation}
\tau =c_s^2\rho _{,T}\delta T+\left[ \frac{\partial p_T}{\partial M^2}+\frac 
12M_T^2\right] M_{,c}^2c_0+\frac 13\left[ \left\langle \omega
^2\right\rangle -M^2\left\langle 1\right\rangle \right] c_0
\end{equation}
Now

\begin{equation}
\frac{\partial p_T}{\partial M^2}=\frac{-1}2M_T^2
\end{equation}
so

\begin{equation}
\tau =-c_0\left\{ \left[ c_s^2-\frac 13\right] \left\langle \omega
^2\right\rangle +\left[ \frac{M^2}3-\frac{c_s^2}2TM_{,T}^2\right]
\left\langle 1\right\rangle \right\} ,
\end{equation}
Using Eqs. (\ref{cnot}), (\ref{omega}) and (\ref{omega2}), we get

\begin{equation}
\tau =-\frac{u_{,\lambda }^\lambda \left[ M^2-\frac 12TM_{,T}^2\right] ^2}{%
3T^5\left( \frac{d\rho }{dT}\right) ^2}\frac{\left\{ \left\langle \omega
^3\right\rangle \left\langle 1\right\rangle -\left\langle \omega
^2\right\rangle \left\langle \omega \right\rangle \right\} ^2}{\left|
\left\langle K\left[ 1\right] \right\rangle \right| }  \label{bulk}
\end{equation}

\subsection{Shear stress and viscosity}

The shear stress can be read directly out of the new terms in $T_T^{\mu \nu
} $. In the rest frame, we get

\begin{equation}
\tau ^{ij}=\frac{-1}{bT}\Gamma _{kl}^mH^{kl}\left\langle
p^ip^j,Y_m\right\rangle =\frac{-1}{bT}H_{kl}\left\langle p^ip^j-\frac 13%
\delta ^{ij}\vec p^2,p^kp^l-\frac 13\delta ^{kl}\vec p^2\right\rangle =\frac{%
-R}{bT}H^{ij}
\end{equation}
from which we can read out the shear viscosity

\begin{equation}
\eta =\frac R{bT}  \label{shear}
\end{equation}

To estimate $\eta $, it is enough to keep only the leading (binary
scattering) contributions, so $\eta \sim \lambda ^{-2}.$ On dimensional
grounds, $b\sim T^2$ and $R\sim T^6$, so we recover the usual result, $\eta
\sim T^3/\lambda ^2$.

\subsection{Bulk viscosity}

As expected, things are not so simple with the bulk viscosity. We can read
it out from Eq. (\ref{bulk}) as

\begin{equation}
\zeta =\frac{\left[ M^{2}-\frac{1}{2}TM_{,T}^{2}\right] ^{2}}{3T^{5}\left( 
\frac{d\rho }{dT}\right) ^{2}}\frac{\left\{ \left\langle \omega
^{3}\right\rangle \left\langle 1\right\rangle -\left\langle \omega
^{2}\right\rangle \left\langle \omega \right\rangle \right\} ^{2}}{\left|
\left\langle K\left[ 1\right] \right\rangle \right| }  \label{bulkvis}
\end{equation}
However, in evaluating it we must consider that $\left\langle 1\right\rangle 
$ is logarithmically divergent in the massless limit, so we must correct the
sheer dimensional estimate to $\left\langle 1\right\rangle \sim T^{2}\ln
\left( M/T\right) $. As for the size of $\left| \left\langle K\left[
1\right] \right\rangle \right| $, observe that the integral is dominated by
the Rayleigh - Jeans tail, where $f_{0}\sim T/\omega \gg 1$. Thus $\left|
\left\langle K\left[ 1\right] \right\rangle \right| \sim \lambda
^{4}T^{6}F\left( M^{2}\right) $. Since the overall units are $Mass^{4}$, it
must be $\left| \left\langle K\left[ 1\right] \right\rangle \right| \sim
T^{6}/M^{2}$. For the remaining elements we may use the conventional
estimates $\left\langle \omega ^{3}\right\rangle \sim T^{5}$, $\rho \sim
T^{4}$, and thus obtain

\begin{equation}
\zeta \sim \frac{M^2}{\lambda ^4T^3}\left[ M^2-\frac 12TM_{,T}^2\right]
^2\ln ^2\left( M/T\right)  \label{sjres}
\end{equation}
which reproduces JY's Eq. (5.6) \cite{jeon:1996a}.

In the limit in which the bare mass vanishes, or equivalently in the $%
T\rightarrow \infty $ limit, we may write on dimensional grounds

\begin{equation}
M^2-\frac 12TM_{,T}^2\equiv \frac 12\mu M_{,\mu }^2\sim \lambda M^2
\label{rg}
\end{equation}
and since $M^2\sim \lambda T^2$ itself, Eq. (\ref{sjres}) reduces to $\zeta
\sim \lambda T^3\ln ^2\left( \lambda \right) $, again in agreement with JY 
\cite{jeon:1996a}.\newline

\noindent {\bf Acknowledgments}

This work was reported at the program on Quantum Field Theory of
Non-Equilibrium Processes at INT- University of Washington. We thank the
organizers, J-P Blaizot, R. Pisarsky and L. Yaffe for their invitation, the
INT staff for their hospitality, and colleagues with whom we overlapped
during our stay, P. Danielewicz, E. Iancu, D. Litim, G. Moore, E. Mottola,
S. Mrowczinsky, D. Son and L. Yaffe, for interesting discussions.

EC is supported in part by CONICET, UBA and Fundaci\'on Antorchas. BLH is
supported in part by NSF grant PHY98-00967 and their collaboration is
supported in part by NSF grant INT95-09847.

\section{Appendix}

\subsection{$\Pi _{12}$ is purely imaginary}

To show this, observe that $\Pi _{12}$ is the sum of all 1PI Feynman graphs
with two external vertices, one carrying a $1$ index and the other a $2$
index (this follows from it being the result of opening one $12$ leg in each
2PI vacuum bubble). It can also been represented as

\begin{equation}
\Pi _{12}=\left. \frac{\partial \Gamma _1}{\partial \phi ^1\partial \phi ^2}%
\right| _{\phi =0}
\end{equation}
where $\Gamma _1$is the usual (1PI) effective action, and $\phi ^a$ the
background field. The effective action has the structure

\begin{equation}
\Gamma _1=\frac 12\int dxdy\;\left\{ \left[ \phi \right] \left( x\right)
D\left( x,y\right) \left\{ \phi \right\} \left( y\right) +i\left[ \phi
\right] \left( x\right) N\left( x,y\right) \left[ \phi \right] \left(
y\right) \right\} +O\left( \phi ^3\right)
\end{equation}
where $\left\{ \phi \right\} =\phi ^1+\phi ^2$, $\left[ \phi \right] =\phi
^1-\phi ^2$; both $D$ and $N$ are real, $N$ is even, and $D$ is causal.
Therefore

\begin{equation}
\frac{\partial \Gamma _1}{\partial \phi ^1\left( x\right) }=\frac 12\int
dy\;\left\{ D\left( x,y\right) \left\{ \phi \right\} \left( y\right) +\left[
\phi \right] \left( y\right) D\left( y,x\right) +2iN\left( x,y\right) \left[
\phi \right] \left( y\right) \right\} +O\left( \phi ^2\right)
\end{equation}
and

\begin{equation}
\Pi _{12}=-iN\left( x,y\right) +\frac 12\left[ D\left( x,y\right) -D\left(
y,x\right) \right]
\end{equation}

The real part of $\Pi _{12}$ is odd, and its imaginary part even, which
shows that its Fourier transform is purely imaginary: Write

\begin{equation}
\Pi _{12}\left( x,y\right) =\int \frac{d^4p}{\left( 2\pi \right) ^4}%
\;e^{ip\left( x-y\right) }\Pi _{12}\left( p\right) ,
\end{equation}
then the identity $\Pi _{12}^{*}\left( x,y\right) =-\Pi _{12}\left(
y,x\right) $ becomes indeed $\Pi _{12}^{*}\left( p\right) =-\Pi _{12}\left(
p\right) .$

We may use the same argument to find that $\Pi _{21}\left( x,y\right) =\Pi
_{12}\left( y,x\right) $, so $\Pi _{21}\left( p\right) =\Pi _{12}\left(
-p\right) $ is also imaginary. We also find

\begin{equation}
\Pi _{11}=iN\left( x,y\right) +\frac 12\left[ D\left( x,y\right) +D\left(
y,x\right) \right]
\end{equation}
from where ${\rm Im}\Pi _{11}\left( p\right) =(i/2)(\Pi _{12}+\Pi _{21}).$

\subsection{$\left\langle \omega \right\rangle $ and $\left\langle \omega
^3\right\rangle $}

Our objective is to compute

\[
\left\langle \omega ^3\right\rangle =\frac 1{2\pi ^2}\int_M^\infty d\omega
\;\omega ^3\left[ \omega ^2-M^2\right] ^{1/2}f_0\left( 1+f_0\right) 
\]

\[
\left\langle \omega \right\rangle =\frac 1{2\pi ^2}\int_M^\infty d\omega
\;\omega \left[ \omega ^2-M^2\right] ^{1/2}f_0\left( 1+f_0\right) 
\]
Recall the identity

\begin{equation}
\omega f_0\left( 1+f_0\right) =T^2\frac{\partial f_0}{\partial T}
\label{fluctuation}
\end{equation}
This and Eq. (\ref{thermoid}) may be used to establish the identity

\begin{equation}
\left\langle \omega ^3\right\rangle -M^2\left\langle \omega \right\rangle
=3T\left( p+\rho \right) =3T^2\frac{d\rho }{dT}c_s^2  \label{eqn1}
\end{equation}

Similarly

\begin{equation}
\left\langle \omega ^3\right\rangle -\frac 12TM_{,T}^2\left\langle \omega
\right\rangle =T^2\frac{d\rho }{dT}  \label{eqn2}
\end{equation}
So

\begin{equation}
\left\langle \omega ^3\right\rangle =T^2\frac{d\rho }{dT}\frac{\left[ M^2-%
\frac 32TM_{,T}^2c_s^2\right] }{\left[ M^2-\frac 12TM_{,T}^2\right] };\qquad
\left\langle \omega \right\rangle =T^2\frac{d\rho }{dT}\frac{\left[
1-3c_s^2\right] }{\left[ M^2-\frac 12TM_{,T}^2\right] }  \label{int1}
\end{equation}

\section{Figure Captions}

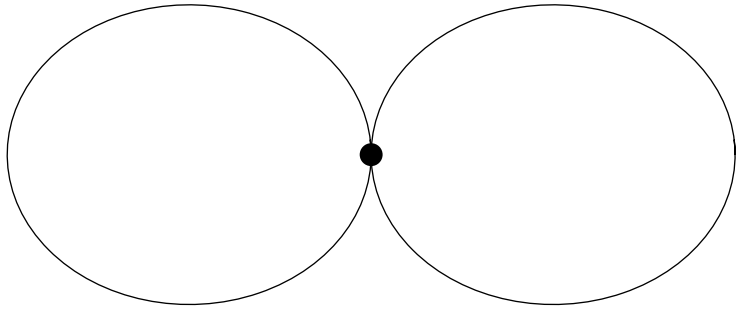
Fig1. Two loops contribution to the CTP effective action.

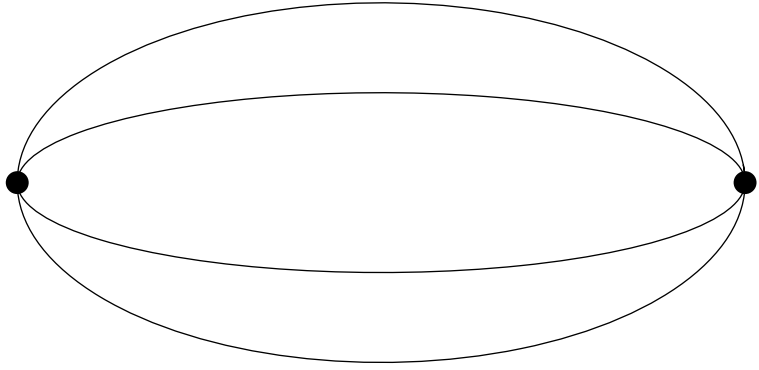
Fig2. Three loops contribution to the CTP effective action.

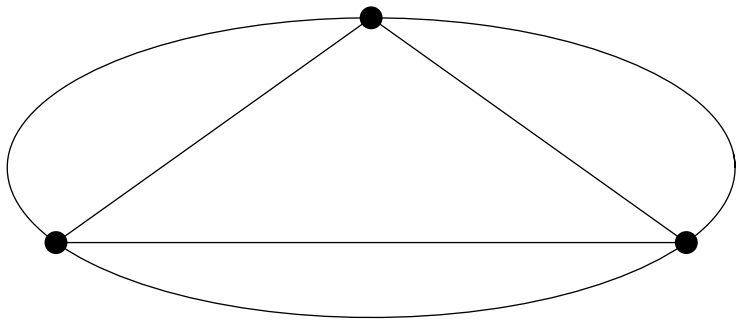
Fig3. Four loops contribution to the CTP effective action.

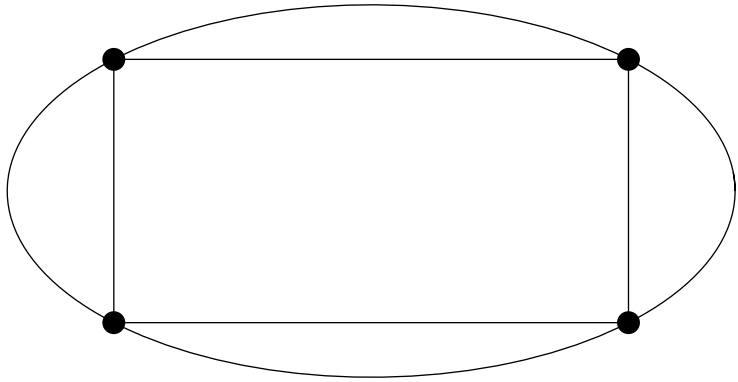
Fig4. Five loops contribution to the CTP effective action.

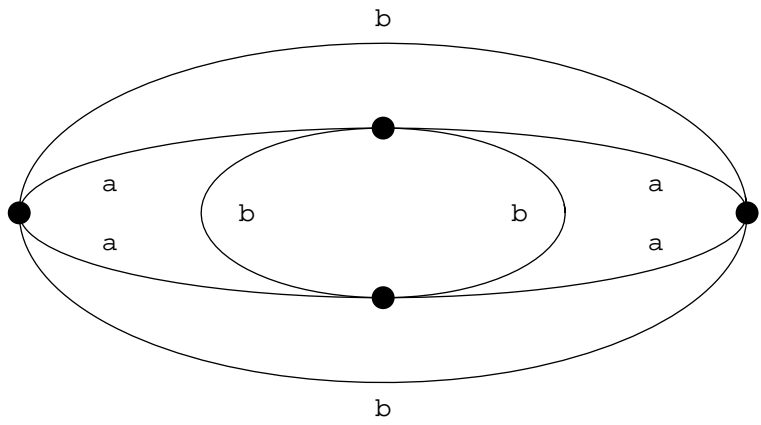
Fig5. The other five loops contribution to the CTP effective action. Observe
the two sets of inequivalent lines, marked {\it a} and {\it b}.

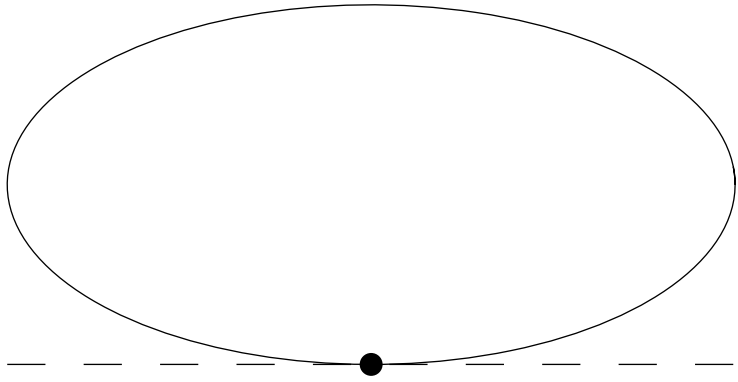
Fig6. One loop contribution to the self energy (tadpole graph)

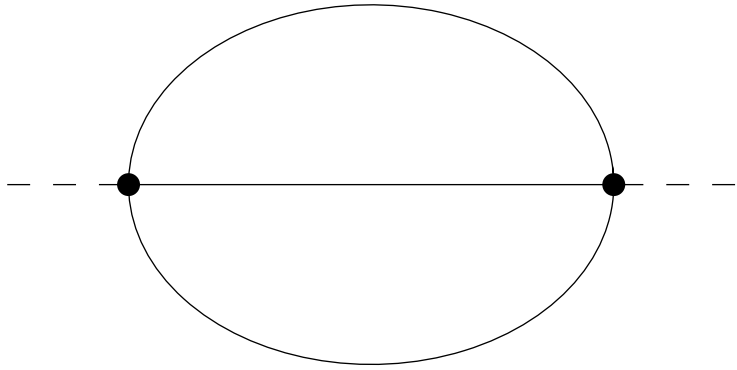
Fig7. Two loops contribution to the self energy (setting sun graph)

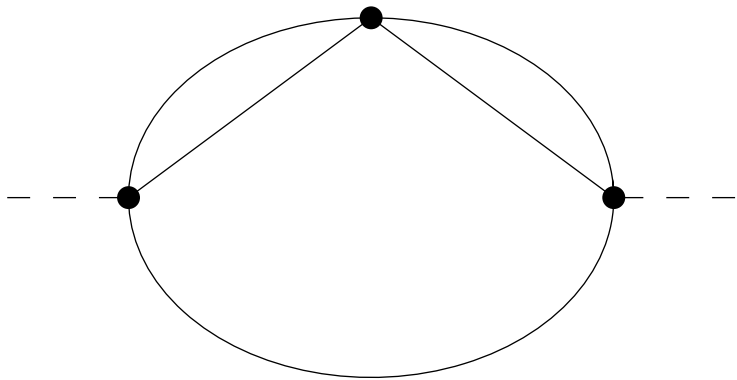
Fig8. Three loops contribution to the self energy.

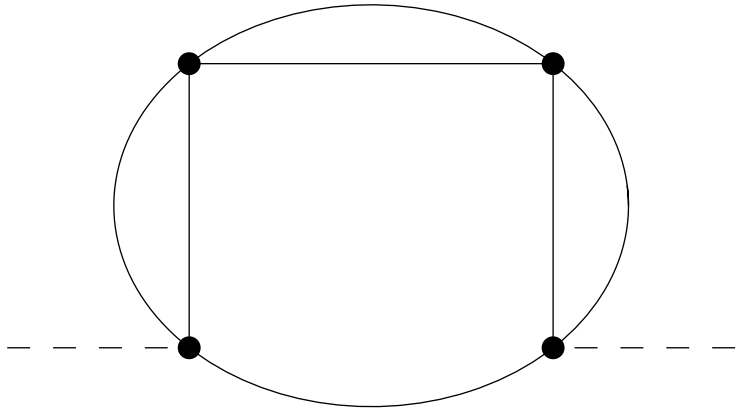
Fig9. Four loops contribution to the self energy

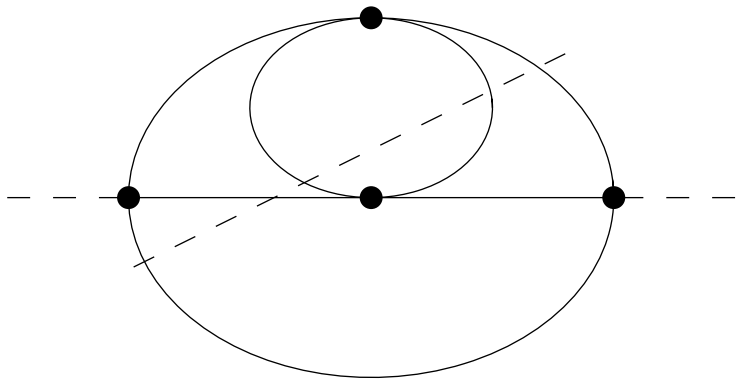
Fig10. Another four loops contribution to the self energy. Cutting as shown,
we go across five internal lines. The symmetric cut also goes across five
lines.

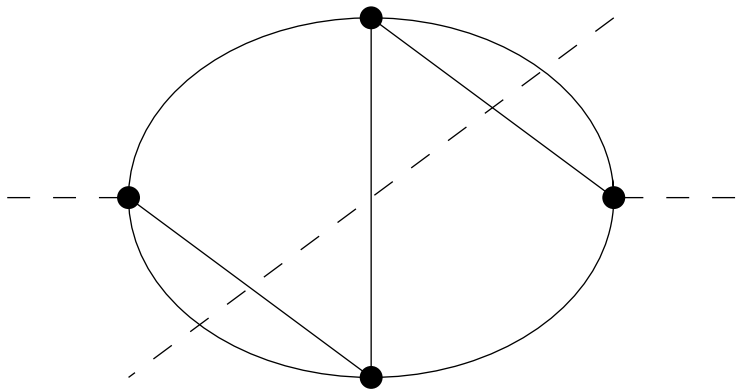
Fig11. The final four loops contribution to the self energy. Cutting as
shown, we go across five internal lines.

\end{document}